\documentclass[aps, preprint, eqsecnum, secnumarabic, nofootinbib,
11pt]{revtex4}

\usepackage{graphicx}
\usepackage{epic}
\usepackage{eepic}
\usepackage{latexsym}
\usepackage{amsfonts}

\newcommand{\eq}[1]{(\ref{#1})}
\newcommand{\be}{\begin{equation}}
\newcommand{\ee}{\end{equation}}
\newcommand{\bea}{\begin{eqnarray}}
\newcommand{\eea}{\end{eqnarray}}
\newcommand{\del}{\partial}
\newcommand{\e}{\epsilon}

\newcommand{\g}{\gamma}
\newcommand{\vs}[1]{\vspace{#1 mm}}
\newcommand{\hs}[1]{\hspace{#1 mm}}

\def\mP{\widehat{\mathcal{P}}}
\def\hM{\widehat{\mathcal{M}}}

\def\ot{\otimes}

\def\a{\alpha}
\def\b{\beta}
\def\c{\gamma}
\def\C{\Gamma}
\def\d{\delta}
\def\D{\Delta}
\def\e{\epsilon}

\def\f{\phi}
\def\fr{\frac}

\def\l{\lambda}
\def\L{\Lambda}
\def\m{\mu}
\def\n{\nu}

\def\s{\sigma}

\def\t{\tau}
\def\th{\theta}
\def\Th{\Theta}

\def\O{\Omega}
\def\x{\xi}

\def\o{\omega}

\def\del{\partial}

\let\bm=\bibitem
\def\nn{\nonumber}

\begin{document}

\baselineskip=.50cm

\rightline{MITF-03-21}
\rightline{UUITP-22/03}

\vs{3}

\title{Multi-Spin Giants}

\author{S. Arapoglu}
\email[e-mail:]{arapoglu@boun.edu.tr}
\affiliation{Bogazici University, Department of
Physics, \\ 34342, Bebek, Istanbul, Turkey}

\author{N.S. Deger}
\email[e-mail:]{deger@gursey.gov.tr}
\author{A. Kaya}
\email[e-mail:]{kaya@gursey.gov.tr}
\affiliation{Feza Gursey Institute,\\
Cengelkoy, 81220, Istanbul, Turkey}

\author{E. Sezgin}
\email[e-mail:]{sezgin@physics.tamu.edu}
\affiliation{George P. \& Cynthia W. Mitchell Institute for
Fundamental Physics,\\ Texas A\&M University, College Station, TX
77843--4242, USA}

\author{P. Sundell}
\email[e-mail:]{per.sundell@teorfys.uu.se}
\affiliation{Uppsala University, Department of Theoretical Physics,\\
Box 803, SE-751 08, Uppsala, Sweden\vs{15}}
%\date{\today}
\begin{abstract}
\baselineskip=.50cm \vs{3}
We examine spherical $p$-branes in $AdS_m\times S^n$, that wrap an
$S^p$ in either $AdS_{m}$ ($p=m-2$) or $S^{n}$ ($p=n-2$). We first
construct a two-spin giant solution expanding in $S^n$ and has spins
both in $AdS_m$ and $S^n$. For $(m,n)=\{(5,5),(4,7),(7,4)\}$, it is
1/2 supersymmetric, and it reduces to the single-spin
giant graviton when the $AdS$ spin vanishes. We study some of its
basic properties such as instantons, non-commutativity, zero-modes,
and the perturbative spectrum. All vibration modes have real and
positive frequencies determined uniquely by the spacetime curvature,
and evenly spaced. We next consider the $(0+1)$-dimensional
sigma-models obtained by keeping generally time-dependent
transverse coordinates, describing warped product of a breathing-mode
and a point-particle on $S^n$ or $AdS_m\times S^1$. The BPS bounds show
that the only spherical supersymmetric solutions are the single
and the two-spin giants. Moreover, we integrate the sigma-model and
separate the canonical variables. We quantize exactly the
point-particle part of the motion, which in local coordinates gives
P\"oschl-Teller type potentials, and calculate its contribution to the
anomalous dimension.
\end{abstract}
\maketitle
\thispagestyle{empty}
\tableofcontents
\baselineskip=.55cm
\section{Introduction}
\setcounter{page}{1} Giant gravitons were first proposed in
\cite{susskind} in order to explain the stringy exclusion
principle \cite{exclusion} where the bound on the R-charge of CFT
operators was related to the bound on the angular momentum in the
supergravity picture. They are probe brane solutions in an $AdS_m \times
S^n$ background with fluxes, obtained by wrapping an ($n-2$)-brane
on an $S^{n-2}$ sphere rotating inside $S^{n}$. Later it was shown
that \cite{myers, hashimoto} an ($m-2$)-brane wrapped on $S^{m-2}$
at constant radius in $AdS_{m}$ and rotating inside $S^n$
carries the same quantum numbers. Together with the Kaluza-Klein
point-like excitation, they constitute three different states
representing the same graviton. However, the states are expected
to be mixed with each other in quantum theory due to the existence
of instantons that would allow semi-classical tunneling
\cite{myers, hashimoto, lee} which may resolve this puzzle.

The giant graviton is an example of how semi-classically stable
string/brane solutions may be helpful in understanding different aspects
of the $AdS$/CFT correspondence. Basically, this is due to the fact that
large charges suppress  quantum fluctuations and thus connect regimes
where both the bulk theory and the CFT have meaningful perturbative
expansions.  From the bulk point of view, the main idea is to zoom in on
a sub-sector of states carrying large charges scaling like the tension of
a $p$-brane, and consider various semi-classical expansion schemes of the
probe brane quantum field theory based on the identification of some
small parameters. Typically, one would expect small parameters to measure
the deviation from being BPS, though interestingly enough there also
exist meaningful expansions in regimes far from being BPS.

This strategy has been successfully utilized in the BMN limit
\cite{BMN} where the relevant states are near BPS and represented as
small closed strings in $AdS_5\times S^5$ with center of mass
rotating around a large circle of $S^5$ with large angular
momentum $J$. One then considers $J\gg 1$ with $\lambda/J^2$ held
fixed, where $\lambda$ is the 't Hooft coupling. The limit
$J\rightarrow \infty$, removes higher order corrections to the
sigma-model leaving the pp-wave geometry, while the SYM side
narrows down to the tower of ``doped'' operators built on top of
the 1/2 BPS single-trace ground state.

Another interesting sector of states that have been studied along
similar lines, have large spin $S$ in $AdS$ \cite{polgub}.
These arise as long rotating strings, corresponding to towers of
single-trace operators doped by derivatives. The rotation induces a strongly
coupled world-sheet sigma model. If $S$ is the only
semi-classically large parameter, the normal-coordinate expansion
gives $1/\sqrt{\lambda}$-corrections to the $AdS$ energy, that are
difficult to match directly with the weakly coupled CFT, though
other qualitative features do match. However, it was discovered
that if one considers states which carry an additional large $S^5$
spin $J$, then the classical $AdS$ energy has a regular expansion
in $\lambda/J^2$. This prompted the proposal that the $AdS$/CFT
duality can be tested in a non-BPS sector by comparing the
$\lambda/J^2$ expansion of the $AdS$ energy obtained from the
classical string sigma-model, with the corresponding quantum
anomalous dimensions in perturbative SYM theory. This has indeed been
supported by recent results in a series of papers (see for
example \cite{tseytlin1}-\cite{tseytlin6})

Quite generally, brane physics exhibits UV/IR mixing in the sense
that energetic branes tend to grow large transverse directions
probing more and more of the background curvature. In the context of
$AdS$/CFT correspondence, this means that already the leading
order of the probe sigma-model expansion share some
qualitative features with the corresponding subset of CFT
operators, most notably the leading linear relation between $AdS$
energy $E$ and other charges. The existence of semi-classically
stable large strings, or other $p$-branes, therefore points to a
sub-sector of (non-BPS) operators with parametrically large bare
dimensions and suppressed anomalous dimensions.

Returning back to giants, they are stabilized by balancing the tension against
electric or magnetic fluxes (the cosmological constants in
$AdS_m\times S^n$), leaving a finite net tension independent of
the $AdS$ scale. Moreover, the string tension runs in $AdS$.
Hence, in the IR limit of the $AdS_m$ the usual flat space
hierarchy is reversed, such that the excitations of the
$(m-2)$-brane field theory become much lighter than the massive
stringy, or M-theoretic, excitations of the brane.

\begin{figure}
\centerline{\includegraphics[width=6.0cm]{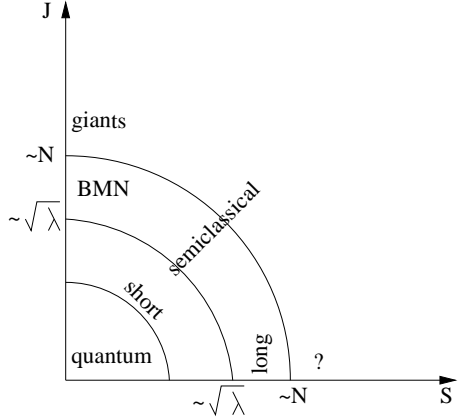}}
\caption{\label{fig1} Different descriptions of the states in
  $AdS_5\times S^5$.}
\end{figure}

In the case of the Type IIB theory on $AdS_5\times S^5$ the
following picture emerges (see figure \ref{fig1}): as the energy
$E$ of string states increases, the flat space Regge trajectories,
where $E$ scales like square-roots of charge, start bending into
linear trajectories as $E\sim {g_sN}$. Roughly speaking, the
energy and some charge of the string concentrate along nearly
light-like portions of the world-sheet, while the transverse
directions can extend resulting for example in the strings on the
pp-wave or the long strings discussed above. At higher energies,
which scale faster with $N$ than the string tension, the
semi-classical string description becomes strongly coupled.
Awaiting some exact world-sheet formulation, the natural
semi-classical probe is instead the $D3$-brane whose tension scales
like $N$. In fact, the tension
is of order $1$ in units of the ten-dimensional Planck length,
which means that the giant $D3$ is the last description prior to
complete breakdown of the geometrical picture in the IIB theory.

The giant picture arises also in M theory on $AdS_{4/7}\times
S^{7/4}$, where $M2$ and $M5$ branes can be dynamically stabilized against
collapse. The semi-classical limits are
essentially the same as for the $D3$-brane, since the tension is
given by the eleven-dimensional Planck length (though here there
is no clear analog on the CFT side of the stringy exclusion
principle). Hence giants appear to capture universal features of
the holography, valid both in string and M-theory and relying only
on the notion of expansion in $1/N$, the bulk Planck's constant.

The above discussion suggests that the appropriate semi-classical
treatment of $p$-brane giants is to expand in $\eta=(E-J)/J$ in
the regime $E\sim J\sim N\gg 1$, $E-J\ll N$. The quantity $E-J$ is
the total energy of the open plus closed string excitations above
the giant ground state, therefore small $\eta$ is the same as considering
a few massless quanta on the giant, described by the $p$-brane
field theory. One may also consider finite values of $\eta$, as
long as one stays safely away from the Planck regime where
$\eta\gg1$. As in the case of the string sigma-model, the problem
of analyzing the normal-coordinate expansion simplifies further in
the double-scaling limit $N\to\infty$, with $\eta$ fixed, where the
$p$-brane field theory reduces to the ground state described by
the classical solution plus the quadratic fluctuations on top of
it. The spectrum of normal frequencies for these vibrations for 1/2
supersymmetric single-spin giant gravitons was calculated in \cite{mathur},
and found to be independent of the size of the brane, and thus the
angular momentum. This implies that in the large $N$ limit of the dual
theory, the corresponding R-charged chiral operator, which is
realized as a sub-determinant \cite{berkooz1, jevicki, vb, berkooz2,
berenstein}, has associated with it a sector of mainly non-BPS
operators with level spacings independent of the ground state
R-charge. In \cite{berenstein} this sector was constructed as
impurities inserted into the ground state sub-determinant mixed
with separate single-traces, shown to produce the structure of a
Fock space of mixed open plus closed string excitations.

In this work, motivated by the fact that semi-classical
string solutions with large $AdS$ spin and its multi-spin
generalizations successfully mimic the BMN strategy, we aim to show
that similar ideas can be extended to giant $p$-branes.
The plan of our paper is as follows.

In section 2, we construct
a 1/2 supersymmetric\footnote{The supersymmetry of the solution
can be established only when there exist a suitable
$\kappa$-symmetric brane action coupled to a supergravity
background. The main examples are $(m,n)=\{(5,5),(4,7),(7,4)\}$.},
spherically symmetric giant $(n-2)$-brane solution in $AdS_m\times
S^n$ that spins both in $AdS_m$ and $S^n$. These two rotations are
rigid and the field equations fix the angular velocities in terms
of the curvature scales, while the spins are fixed by the radii of
the circles of rotation. By adjusting the former radius, we can
take the $AdS$ spin to be small or large. We show the saturation
of a BPS bound, where the energy is equal to the sum of the
two angular momenta. There is also a point particle limit with
non-vanishing conserved quantities, connected to the giant by an
instanton (with finite action). In \cite{das}, it was suggested
that imposing supersymmetry causes non-commutativity in the phase
space. We indeed observe this to happen also in the two-spin
case, with additional Dirac brackets between the
radial $AdS$ coordinate and the two cyclic coordinates in $AdS_m$
and $S^n$ used for the rotations.

In Section 3, we examine various aspects of the vibration spectrum
of the single and the two-spin BPS giants. The bosonic fluctuations of the
two-spin BPS giant (we are only considering the scalar fluctuations and
leave vector and tensor fluctuations on $D3$ and $M5$ branes for
future work) has two interesting features. Firstly, the
frequencies depend only on the curvature scales of $AdS_m$ and
$S^n$, despite the fact that there are three more length scales,
namely the radius of the rotation in $AdS_m$, the tension and the size of
the brane. As a consequence, the vibration spectrum is evenly
spaced, which is in agreement with the large-$N$ Fock space
picture of \cite{berenstein}. Secondly, for generic $m$ and $n$,
the vibration spectrum of the corresponding single-spin giant
graviton is contained as a subset in the spectrum of two-spin
giant. In fact, when $(m,n)=\{(5,5),(4,7),(7,4)\}$, the two
spectra become identical but with different degeneracy. We also
work out the fermionic vibration modes for single and two-spin BPS giant
$M2$ in $AdS_7\times S^4$ background.

In Section 4 we construct more general spherical giants where all
available coordinates are assumed to depend on time after
identifying the brane directions in space-time. They can
support maximum possible number of independent spins both for
the branes expanding in $AdS$ and on the sphere. We find that the
truncated $p$-brane equations can be integrated, and the canonical
variables become separated, leading to an interesting set of
potentials both for the ``breathing mode'' and the remaining
``point-particle'' motion. In fact, the latter we find to be
governed by trigonometric and hyperbolic generalizations of the Calogero
model, known as P\"oschl-Teller Type I and II potentials, which are
exactly solvable quantum mechanics models (see \cite{cooper} and
\cite{ant} for review). Alternatively, the point-particle sector can be
quantized using global coordinates on embedding space leading to ordinary
spherical harmonics. We also derive BPS bounds on the energy, and show
that they can be saturated only by the single-spin and the two-spin
solutions found in this paper, which are hence the only BPS spherical
giants. Finally, we quantize the $(0+1)$-dimensional sigma-model,
treating the point-particle motion exactly while borrowing the results
from \cite{ouyang} for the breathing mode obtained using the
Bohr-Sommerfeld recipe \cite{min}. Interestingly enough, the exact energy
spectrum from the (in general non-supersymmetric) point-particle motion
is evenly spaced, while the breathing gives complicated corrections to
the energy. This leads to a prediction for anomalous dimensions, however
it is difficult to pinpoint the operators precisely.

In Section 5 we conclude and discuss several open problems. The
Appendices contain details of the methods used to quantize the
point-particle sector of the 0+1 dimensional sigma model, namely the
global coordinates formulation, the eigenvalue spectra of P\"oschl-Teller
Type I and II potentials, and a comparison of the exact results with the
Bohr-Sommerfeld method.

\vs{2}

\hs{-5}{\bf Note Added:} In an earlier version of this paper, solutions
of \eq{get} which enhance 1/4 supersymmetry to 1/2 were overlooked and
the two-spin giant solution presented in section 2 was erroneously
claimed to be 1/4 supersymmetric. This mistake was corrected after the
appearance of \cite{mp} where it was shown that the two-spin solution is
related to the single-spin solution by an $AdS$ isometry  and thus has 1/2
supersymmetry.

\section{1/2 Supersymmetric Two-Spin Giants}

In this section we construct a 1/2 supersymmetric two-spin giant
graviton wrapped on $S^{n-2}$ inside $S^n$ of $AdS_m\times S^n$
and rotating simultaneously on $S^n$ and $AdS_m$. The bosonic
$p$-brane action can be written as \be S=-T_p \int d^{p+1}\sigma
\sqrt{-\g}\,\,\left[
1+\frac{1}{(p+1)!}\,\,\e^{\a_0...\a_p}\del_{\a_0}X^{M_0}..\del_{\a_p}X^{M_p}
A_{M_0..M_p}\right]\label{act} \ee
where $\g_{\a\b}$ is the pull-back of the space-time metric to the
world-volume. In some cases like $M5$ or $D3$-branes, there are
additional world-volume fields which can be consistently set to
zero. The field equations of the above action are
\bea
\frac{1}{\sqrt{-\g}}\partial_\a\left[\sqrt{-\g}\g^{\a\b}\partial_\b
  X^N g_{MN}\right]
-\frac{1}{2}\g^{\a\b}\partial_\a X^N\partial_\b X^P
\partial_M g_{NP}=\nonumber\\
\frac{1}{(p+1)!}\,\,\e^{\a_0...\a_p}\,
\partial_{\a_0}X^{M_0}...\partial_{\a_p}X^{M_p}
H_{MM_0...M_{p}},\label{field} \eea
where $H$ is the field strength of the $(p+1)$-form potential,
i.e. $H=dA$. The metric of $AdS_m \times S^n$ is:
\be\label{metrik1} ds^2=-f dt^2+ f^{-1}\,dr^2+ r^2 d\O_{m-2}^2+
L^2 d\O_n^2 \ee where $f=1+r^2/\tilde{L}^2$ and \bea d\O_n^2 &=&
d\th^2 +{\rm cos}^2\th d\f^2+{\rm sin}^2\th\left[d\chi_1 ^2+ {\rm
sin}^2\chi_1 (...+ {\rm sin}^2\chi_{n-3}d\chi_{n-2} ^2)\right]\
\label{sp1}
,\\
d\O_{m-2}^2 &=& d\a_1^2 + {\rm sin}^2\a_1\left[d\a_2^2 +{\rm
sin}^2\a_2 (...+{\rm sin}^2\a_{m-3}d\a_{m-2}^2)\right].
\label{sphere} \eea
Here $L$ and $\tilde{L}$ are the radius of curvatures which are
related as $(m-1)L=(n-1)\tilde{L}$. When appropriate form fields
are turned on which have non-zero fluxes on $AdS$ or on the
sphere, these geometries compromise maximally supersymmetric
backgrounds of the corresponding supergravities for certain values
of $m$ and $n$. In this section we are interested in the magnetic
backgrounds and thus should turn on the flux on the sphere. In the
above coordinates the appropriate ($n-1$)-form potential
supporting this flux becomes
$A_{\phi\chi_1..\chi_{n-2}}=L^{n-1}(\sin\theta)^{n-1}
\sqrt{g^\chi}$, where $(g^\chi)_{ij}$ is the metric on the unit
$S^{n-2}$ in (\ref{sp1}) parametrized by $\chi$ coordinates.
\footnote{In some cases, such as $D3$, the $n$-form flux is
self-dual and the potential has also an electric part. However,
this does not affect the field equations for our ansatz. This
remark is valid for all solutions found in this paper. \label{ft}}

It is easy to verify that the following configuration solves the
field equations (\ref{field}): \bea
&&t=\tau,\,\,\,\,\, \chi^i=\sigma^i, \nonumber\\
&&\phi=\tau/L, \,\,\,\,\,\a\equiv\a_{m-2}=\tau/\tilde{L}, \label{sol}\\
&&\theta=\theta_0, \,\,\,r=r_0,
\,\,\,\a_1=..=\a_{m-3}=\pi/2\nonumber \eea
The brane wraps an ($n-2$)-sphere in $S^n$ and rotates both on
$S^n$ and in $AdS_m$ with constant angular velocities which are
fixed by the corresponding curvature scales.

\subsection{BPS Bound}

From the action (\ref{act}), the Lagrangian with $\a=\a(\tau)$ and
$\f=\f(\tau)$  can be obtained as \be {\cal
L}=\fr{N}{L}\left[-(\sin\th)^{n-2}\Delta
+L(\sin\th)^{n-1}\dot{\f}\right], \ee where
$\Delta^2=f-r^2\dot{\a}^2-L^2(\cos\theta)^2\dot{\phi}^2$. Here we
have used the flux quantization $T_{p}A_{p}=N/L^{p+1}$ where $A_p$
is the area of the unit $p$-sphere. The conserved angular momenta
become \bea P_\phi &=&
\frac{NL(\sin\theta)^{n-2}(\cos\theta)^2\dot{\phi}}{\Delta}
+ N (\sin\theta)^{n-1},\label{m1}\\
P_{\a}&=&\frac{N r^2 (\sin\theta)^{n-2}\dot{\a}}{L
\Delta},\label{m2} \eea
and the Hamiltonian can be written as \be
H=P_\f\dot{\f}+P_\a\dot{\a}-{\cal L}=\frac{N (\sin\theta)^{n-2}
  f}{L\Delta}.
\ee
The gauge condition $t=\tau$ implies $H=-P_t=E$, where $E$ is the
conserved $AdS$ energy. From (\ref{m1}) and (\ref{m2}) it is
possible to express $\Delta$ in terms of $P_\phi$ and $P_\a$ which
gives the Hamiltonian as \be\label{ham1}
H=\sqrt{f}\left[\fr{P_\a^2}{r^2}+\fr{P_\phi^2}{L^2}+
  \fr{N^2}{L^2}\tan^2\th(\fr{P_\phi}{N}-\sin^{n-3}\th)^2\right]^{1/2}
\ee
We see from (\ref{ham1}) that the Hamiltonian obeys \be\label{b1}
H\geq \sqrt{f}\,\sqrt{\fr{P_\a^2}{r^2}+\fr{P_\phi^2}{L^2}}
=\sqrt{\left[\fr{P_\phi}{L}+\fr{P_\a}{\tilde{L}}\right]^2+\left[\fr{P_\phi
    r}{L\tilde{L}}-\fr{P_\a}{r}\right]^2}\, ,
\ee
which implies the BPS bound \be H\geq
\fr{P_\phi}{L}+\fr{P_\a}{\tilde{L}}\, .\label{bps2} \ee
For fixed angular momenta, the extremum of the Hamiltonian can be
found from $\del_r H=0$ and $\del_\th H=0$. In the equilibrium $r$
is uniquely determined by $r_0^2= P_\a L\tilde{L}/P_\phi$. On the
other hand, the roots of $\th$ turn out to be equal to the ones
found in \cite{myers}, which are at $\th=0$ and $(\sin\th)^{n-3}=
P_\phi/N$. For the solution (\ref{sol}), the conserved quantities
are
\bea
P_\phi&=&N(\sin\theta_0)^{n-3},\label{mphi}\\
P_\a&=&\frac{Nr_0^2}{L\tilde{L}}(\sin\theta_0)^{n-3},\label{ma}\\
H&=&\frac{P_\phi}{L}+\frac{P_\a}{\tilde{L}}.\label{bps} \eea
As for ordinary giants, $P_\phi$ depends only on the size of the
brane. On the other hand, $P_\a$ is fixed both by the size of the
brane and the radius of the rotation in $AdS$ space.  Note that
$P_\f\leq N$ but $P_\a$ is unbounded. Also, as $r_0\to0$,\,\,we
have $P_\a\to0$ and the single spin solution is recovered.

\subsection{Supersymmetry}
\label{susysection}

Eq. \eq{bps} shows that the energy saturates the BPS bound
(\ref{bps2}) and thus one expects this configuration to be
supersymmetric. We will demonstrate this explicitly for a giant
$M2$ brane in $AdS_7\times S^4$. For other cases when
$(m,n)=\{(5,5),(4,7)\}$ the calculation is similar. As discussed in
\cite{duff}, this solution will have residual supersymmetry if the
following constraint is satisfied
\be \Gamma \e =\e,\label{kil1} \ee where \be
\C=-\frac{1}{3!}\,\,\e^{\a_0\a_1\a_2}\del_{\a_0}X^{M}\del_{\a_1}X^{N}
\del_{\a_2}X^{P} \C_{MNP}, \ee $\e=\e(X)|_{M2}$ and $\e(X)$ is the
Killing spinor in $AdS_7\times S^4$ which can be found explicitly
as (see, e.g., \cite{myers})\footnote{Note the sign differences with
  \cite{myers} in some exponentials.}
\bea \e(X)&=&e^{\fr12 \th \c
  \C^\th}e^{\fr12\phi\c\C^\phi}e^{-\fr12\chi_1\C^{\chi_1\th}}
e^{-\fr12\chi_2\C^{\chi_2\chi_1}}\nonumber\\
&&e^{\fr12 u\C^{r}\c} e^{-\fr{t}{2\tilde{L}}\C^{t}\c}e^{-\fr12
  \a_1\C^{\a_1r}} e^{-\fr12\a_2\C^{\a_2\a_1}}...\,\,e^{-\fr12 \a_5
  \C^{\a_5\a_4}} \e_0, \label{kilsip}
\eea where \bea \sinh
u=r/\tilde{L},\hs{8}\c=\C^{\th\phi\chi_1\chi_2}, \eea $\e_0$ is a
constant spinor and the indices on the gamma matrices refer to the
tangent space. For this background we have $\tilde{L}=2L$.
For the solution (\ref{sol}), eq. (\ref{kil1}) can be
written as (after multiplying by $\C^{\phi\chi_1\chi_2}$ from the
left) \be \left[\C^{t\phi}\cosh u -\C^{\a_5\phi}\sinh u
  -\c\C^{\th}\sin\th + \cos\th \right]\e=0.\label{kil2}
\ee The last two terms can be grouped as $\exp[-\th \c\C^{\th}]$.
Using this and commuting other factors with the first line of
(\ref{kilsip}) we find that (\ref{kil2}) is equivalent to
\be\label{20} \left[\C^{t\phi}\cosh u -\C^{\a_5\phi}\sinh
u+I\right] e^{\fr12 u\C^{r}\c} e^{-\fr{\tau}{2\tilde{L}}\C^{t}\c}
e^{-\fr12
  \a_1\C^{\a_1r}} e^{-\fr12\a_2\C^{\a_2\a_1}}...\,\,e^{-\fr{\tau}{2\tilde{L}}
  \C^{\a_5\a_4}} \e_0=0.
\ee Multiplying from the left by $\exp[\fr12 u \C^{r}\c]$ one
obtains \bea \left[\C^{t\phi}\cosh u -\C^{\a_5\phi}\sinh u+ \cosh
u +
  \C^{r}\c\sinh u \right] e^{-\fr12
  \a_1\C^{\a_1r}} e^{-\fr12\a_2\C^{\a_2\a_1}}...\,\,e^{-\fr{\tau}{2\tilde{L}}
  (\C^{\a_5\a_4}+\C^{t}\c)} \e_0=0,\nn
\eea where we have also carried
$\exp[-\fr{\tau}{2\tilde{L}}\C^{t}\c ]$ to the right. To take care
of the angular dependencies in the middle, we first multiply the
above expression from the left by $\exp[\fr12 \a_1
  \C^{\a_1 r}]$ which commutes with the first three terms and
anti-commutes with the fourth one. This last term gives the
combination \be (\sinh u)\C^{r}\,\c\, e^{-\a_1 \C^{\a_1
r}}=-(\sinh u)\C^{r}\,\c\,\C^{\a_1
  r}=(\sinh u)\, \c\,\C^{\a_1}
\ee
where we have used the fact that $\a_1=\pi/2$. Carrying out
the same calculation for $\a_2,\a_3$ and $\a_4$ we finally get
\be\label{get}
\left[\C^{t\phi}\cosh u -\C^{\a_5\phi}\sinh u+
\cosh u +
  \C^{\a_4}\c\sinh u \right] \,\,e^{-\fr{\tau}{2\tilde{L}}
  (\C^{\a_5\a_4}+\C^{t}\c)} \e_0=0.
\ee
For this equation to hold, $\e_0$ should obey
\bea
&&\left[I+\C^{t\phi}-\tanh u\,
\C^{\a_5\phi}(I+\C^{\a_5\a_4}\c\C^{\phi})\right]\e_0=0, \label{dogru1}\\
&&\left[I+\C^{t\phi}+\tanh u\,
\C^{\a_5\phi}(I+\C^{\a_5\a_4}\c\C^{\phi})\right]
\left[I-\C^{\a_5\a_4t}\c\right] \e_0=0,\label{dogru2} \eea where the
first and the second conditions are implied by the even and the odd
powers of $\tau$ in \eq{get}, respectively. Decomposing $\e_0$ as \be
\e_0=\e^{++}+\e^{+-}+\e^{-+}+\e^{--}, \ee where \be
\C^{t\phi}\,\e^{s_1s_2}\,=\,s_1\,\e^{s_1 s_2},\hs{6}
\C^{\a_5\a_4}\c\C^{\phi}\,
\e^{s_1s_2}\,=\,s_2\,\e^{s_1s_2},\hs{6}s_{1,2}=\pm, \ee one finds that
\eq{dogru1} gives \be \e^{++}=0,\hs{6} \e^{+-}=\tanh u
\,\C^{\a_5\phi}\e^{-+}, \ee and \eq{dogru2} is satisfied identically.
Thus the two-spin giant configuration preserves 1/2 of the
supersymmetries of eleven dimensional supergravity.  The single-spin
solution can be recovered by letting $u\to0$ (i.e. $r\to0$) so that the
motion in $\a_5$ disappears. In that case, the Killing spinor condition
\eq{dogru1} becomes \be (I+\C^{t\phi})\e_0=0, \label{cond1/2} \ee which
is the projection found in \cite{myers,hashimoto}. Indeed, even for
non-zero $u$ it is possible to obtain \eq{cond1/2} from \eq{dogru1} by a
similarity transformation
\be \label{similarity} \e_0\,\to\,{\mathbb S}^{-1}\,
\e_0\hs{10}\C^{A}\,\to\, {\mathbb S}^{-1}\, \C^A\, {\mathbb S} \ee
where
\be
{\mathbb S}=\exp(-\frac12 u\,\C^{t\a_5})\exp(\pm\frac12 \delta\,
\C^{\phi\a_1\a_2\a_3\a_5r})\ ,\qquad \cos \delta=1/\cosh u\ ,
\ee
and the sign in ${\mathbb S}$ is correlated with $\C^{tr\a_1..\a_5}\c=\pm
I$.

\subsection{Instantons}

From the experience with the single spin giants, it is plausible
to expect a point particle limit. Indeed, taking $\sin\th=\e$,\,\,
$L^2 \dot{\phi}^2=(1-\e^{2n-4})/(1-\e^2)$ and letting $\e\to0$, we
get finite $P_\phi$ and $P_\a$. Quantum mechanically, there may
exist tunneling between the expanded and the point-like
configurations. Following \cite{myers}, to construct the relevant
instanton solution we first assume $\th=\th(\tau)$ and let $\tau
\to i z$. Using the corresponding conserved Euclidean energy we
get a first order equation \be
P_\phi\,L\,\fr{d\th}{dz}=\tan\th[P_\phi-N(\sin\th)^{n-3}]\, , \ee
which is  exactly the same one obtained in \cite{myers}. The
solution is \be
(\sin\th)^{n-3}=\fr{P_\phi\,\,e^{(n-3)z/L}}{1+\,N\,\,e^{(n-3)z/L}}
\ee which interpolates between the extrema of $\th$ as
$z\to\pm\infty$. The total Euclidean action for this instanton is
finite and thus one would expect mixing between the expanded and
zero-size branes  by tunneling. Note that unlike the ordinary
giants, here there is no dual 1/2 BPS spherical configuration
corresponding to a  brane expanding in $AdS$. Because of this, no
puzzle arises in solving the stringy exclusion principle.

\subsection{Non-commutativity}

In \cite{das}, a relation between supersymmetry and
non-commutativity in the phase space was proposed, i.e. the BPS
condition gives constraints and Dirac type canonical quantization
leads to non-commutativity. One may wonder the consequences of
having spin in $AdS$, as in our solution, for this analysis. For
that purpose we relax the conditions on $r$ and $\th$ coordinates in
\eq{sol} and let them to be dynamical. This modifies the
Hamiltonian (\ref{ham1}) so that there are additional $fP_r^2$ and
$P_\th^2/L^2$  terms inside the square-root. For our 1/2
supersymmetric two-spin configuration we now have two primary constraints
\be \psi_1\equiv P_\th=0,\,\,\,\,\,\,\psi_2\equiv P_r=0. \ee The
Poisson brackets of these with the Hamiltonian give two secondary
constraints  which are \be
\psi_3=\fr{dV}{d\th},\,\,\,\,\,\,\psi_4=\fr{r^4}{\tilde{L}^2}
\left[V(\th)+\fr{P_\f^2}{L^2}\right]-P_\a^2, \ee where
$V(\th)=N^2\tan^2\th(P_\f/N-\sin^{n-3}\th)^2/L^2$. There are no
further constraints since the Poisson brackets become
$\{H,\psi_3\}\sim P_\th$ and $\{H,\psi_4\}\sim P_r$. The
constraints are second class with the algebra \be
\{\psi_1,\psi_3\}=-\fr{d^2V}{d\th^2},\,\,\,\,\,\,
\{\psi_2,\psi_4\}=-\fr{4P_\a^2}{r}. \ee We also have
$\{\psi_1,\psi_4\}\sim\psi_3$ which vanishes on the restricted
surface. The canonical structure on this constrained phase space
can now be described by defining a Dirac bracket as follows:
$\{f,g\}_{DB}\equiv\{f,g\}-\{f,\psi_i\}(C^{-1})^{ij}\{g,\psi_j\}$
where $C_{ij}\equiv\{\psi_i,\psi_j\}$. This gives
\bea
\{\th,\f\}_{DB}& =& \fr{1}{N(n-3)(\sin\th)^{n-4}\cos\th}\, , \nn \\
\{r ,\a\}_{DB} &=& \fr{L\tilde{L}}{2Nr(\sin\th)^{n-3}}\, ,\\
\{r,\phi\}_{DB}&=& \fr{rL^2\tilde{L}^2}{2N(\sin\th)^{n-3}}\, .
\nn
\eea
As in \cite{das}, non-commutativity is proportional to $1/N$.
Note that it occurs not only between the sphere coordinates $\th$
and $\f$ but also between $AdS$ coordinates $r$ and $\a$ and
between $r$ and $\f$. For the single spin case, $P_\a=0$ is an
extra primary constraint due to which $\{\psi_2,\psi_4\}=0$ and
non-commutativity exists only between $\th$ and $\f$ \cite{das}.

\subsection{Validity of the Solution}
\label{val} One can trust the solution only when the corrections
to the space-time geometry and the Born-Infeld action can be
ignored. In string theory the former requires $g_s\ll 1$ and
$g_s\,N\gg1$. On the other hand, the corrections to the
Born-Infeld action are suppressed if the induced curvature scale
on the world-volume is much larger than the string scale which
gives $L\sin\th_0\gg \sqrt{\a'}$. For $AdS_5\times S^5$, $L^4=4\pi
g_s N\a'^2$ and thus we have $(Ng_s)^{1/4}\sin\th_0\gg 1$. This
implies from (\ref{mphi}) $P_\f\gg \sqrt{N/g_s}$. In M-theory, we
still need $N\gg 1$ to rely on background geometry. To remove
higher derivative corrections to $M2$ and $M5$ brane actions it is
necessary to have $L\sin\th_0\gg l_p$. For $M2$ \,$L\sim l_p
N^{1/6}$ and for $M5$\, $L\sim l_p N^{1/3}$ which yields $P_\f\gg
N^{2/3}$ and $P_\f\gg N^{1/3}$ respectively. Therefore, the giant
graviton picture is reliable only when $P_\f$ is large.

\subsection{The Algebra of Unbroken Symmetries} \label{alg}

The bosonic field theory defined by the action (\ref{act}) in
$AdS_m\times S^n$ background carries a representation of the isometry
group  $SO(2,m-1)\times SO(n+1)$, generated by the Killing vectors,
$\delta X^M=K^M(X^N)$.  The Noether charges are denoted by
\bea &&(\hM_{ab},\mP_{a})\ ,\qquad a,b=(t,r,\a,\a_{m-3},\a_s)\ ,\quad
s=1,..,m-4\ ,\label{hatMab}\\ &&(\hM_{IJ},\mP_I)\ ,\qquad
I,J=(\th,\f,\chi^i)\ ,\quad i=1,\dots,n-2\ ,\label{hatMIJ}\eea
where the indices are flat. The generators are anti-hermitian which are
normalized such that
\bea &&{[}\hM_{ab},\hM^{cd}{]}=4\delta_{[a}^{[c}\hM_{b]}{}^{d]}\
,\qquad {[}\hM_{IJ},\hM^{KL]}=4\delta_{[I}^{[K}\hM_{J]}{}^{L]}\ ,\nn\\
&& {[}\hM_{ab},\mP^c{]}=2\delta_{[a}^c\mP_{b]}\ ,\hs{16}
{[}\hM_{IJ},\mP^K]=2\delta_{[I}^K\mP_{J]}\ ,\\
&&{[}\mP_a,\mP_b{]}=-\hM_{ab}/\tilde L^2\ ,\hs{14}
{[}\mP_I,\mP_J{]}=\hM_{IJ}.\nn \eea
In determining the supersymmetry algebra in the background specified by a
solution, the anti-commutator of two unbroken supersymmetries closes on
unbroken bosonic symmetries. For both single and two-spin solutions these
symmetries are
\be U(1)_{H'}\times SO(m-1) \times SO(n-1)_\chi\ ,\label{unbroken}\ee
where $H'$ is the generator of $\tau$-translations and $SO(n-1)_\chi$ are
rotations of the $S^p$ ($p=n-2$) in the $p$-brane worldvolume, i.e.
\bea 
H'&=& \left\{\begin{array}{ll}E-J/L-S/\tilde L&
\mbox{two-spin}\\E-J/L\qquad\qquad&\mbox{single-spin}\end{array}\right.\label{Htwo}\\
SO(n-1)_\chi&:&\quad (\hM_{\th i},\hM_{ij})\ ,\qquad i=1,\dots,n-1\ ,\eea
where
\be E\equiv -i\mP_t\ ,\qquad S\equiv i\hM_{\a\a_{m-3}}\ ,\qquad J\equiv
i\mP_\f\ .\label{U1charges}\ee
The form of $H'$ follows from that $t=\tau=\phi/L=\alpha/\tilde L$ in the
case of the two-spin solution, and $t=\tau=\phi/L$ in the case of the
single-spin solution. In the linearized theory, $H'$ becomes the
fluctuation Hamiltonian. The unbroken $SO(m-1)$ is an internal symmetry
from the point of view of the $p$-brane worldvolume theory. In the case
of single-spin all $SO(m-1)$ rotations are manifest, and its generators
are
\be SO(m-1)\,:\quad {\cal M}_{I'J'}=({\cal M}_{r\a_r},{\cal
M}_{\a_r\a_s})\ ,\qquad I'=1,...,m-1\ ,\quad r=1,\dots,m-2\
,\label{mijprime}\ee
In the case of two-spin an $SO(m-3)\subset SO(m-1)$ subgroup is manifest,
and generated by the rotations preserving the equator of $S^{m-2}\subset
S^m$, i.e.
\be SO(m-3)\,:\quad (\hM_{r\a_r},\hM_{\a_r\a_s})\ ,\qquad r=1,\dots,m-4\
.\ee
In addition to the unbroken symmetries, the linearized $p$-brane action
is invariant under separate shifts in the cyclic $t$, $\alpha$ and $\phi$
coordinates generated by $E$, $S$ and $J$ defined in \eq{U1charges}. Of
these isometries only the combined shift generated by $H'$ given in
\eq{Htwo} is an unbroken isometry. However, the cyclic coordinates are axionic
fields on the $p$-brane and hence $E$, $S$ and $J$ remain
conserved in the linearized theory. These charges generate outer
automorphisms of the the supersymmetry algebra in the background of the
solution. We determine the $U(1)$-charges of the unbroken supersymmetry
charges and fluctuation fields in Section 3.6 in the case of the
single-spin solution.

As an example, which will be used in Section 3.6, let us determine the
unbroken superalgebra in presence of a single-spin giant $M2$-brane in
$AdS_7\times S^4$. The supersymmetry algebra of M-theory expanded around
a solution with non-trivial four-form fluxes is given in \cite{dewit} and
on $AdS_7\times S^4$ it becomes ($\tilde{L}=2L$)
\bea &&\{\widehat{\cal Q},\widehat{\bar{\cal Q}}\}=-2 \C^a \,\mP_a +
\fr{1}{\tilde{L}}\,\C^{ab} \,\c \,\hM_{ab}
-\fr{2}{L} \,\C^I \,\mP_I - \fr{1}{L}\, \C^{IJ}\,\c\,\hM_{IJ},\nn\\
&&{[}\mP_a,\widehat{\bar{\cal Q}}{]}=\fr{1}{2\tilde{L}}\widehat{\bar{\cal
Q}}\C_a\c,
\hs{8}{[}\hM_{ab},\widehat{\bar{\cal Q}}{]}=\fr{1}{2} \widehat{\bar{\cal Q}} \C_{ab},\\
&&{[}\mP_I,\widehat{\bar{\cal Q}}{]}=-\fr{1}{2}\widehat{\bar{\cal
Q}}\C_I\c,\hs{8}{[}\hM_{IJ},\widehat{\bar{\cal Q}}{]}=
\fr{1}{2}\widehat{\bar{\cal Q}}\C_{IJ}.\nn \eea
The full $M2$ brane supersymmetry generators are given by \be
\label{sc}\widehat{\bar{\cal Q}} \e_0=\int d^2\sigma
\sqrt{-\c}\,\gamma^{0i}\,\bar \epsilon (1-\C)\Gamma_i\Theta\ ,
\label{M2fermicharge}\ee
where $\e=g(X)\e_0$ is the Killing spinor \eq{kilsip}.

From \eq{cond1/2}, the unbroken supersymmetry charges for the single-spin
solution are given by \be Q=P_+\widehat{{\cal Q}}\ ,\qquad
P_+=\fr{1}{2}\,(1+\C^{t\phi}). \ee
These obey \be \{Q,\bar{Q}\} = 2P_{+}\left(i\C^t\,H' + \fr{1}{2 \tilde
L}\,\C^{I'J'}\,\c\,{\cal M}_{I'J'}-\fr{1}{2L}\C^{i'j'}\c\,{\cal M}_{i'j'}
\right)\ ,\label{q2}\label{qq1/2}\ee and $[\tilde L E,\bar Q]=[J,\bar
Q]=-{i\over 2}\bar Q\C^t\c$. Next, we define
\be Q^\pm=\Pi_\pm Q\ ,\qquad \Pi_\pm={1\over 2}(1\pm i\C^t \c)\ ,\ee
which commute with $P_{+}$ and the unbroken bosonic symmetries. Hence \be
[H',Q^\pm]=\mp {1\over 4L}Q^\pm\ ,\qquad [J,Q^\pm]=\pm{1\over 2}\
,\label{susychargesingle} \ee and $Q^+$ and $Q^-$ transform as $(4,2)$
and $({\bar 4},2)$ under $SO(6)\times SO(3)_\chi$ which we denote them by
$Q_\a^A$ and $\bar Q^\a_A$, respectively, with\footnote{We use the
following conventions: hermitian generators $L_i$ of $SO(3)$ and Pauli
matrices obey $[L_i,L_j]=i\e_{ijk}L_k$ and
$\s^i\s^j=i\e^{ijk}\s^k+\d^{ij}$. Doublet indices are contracted using
$\e_{AB}=(\e^{AB})^\dagger$, $\e_{AB}\e^{CD}=2\d_{AB}^{CD}$. The
symmetric Pauli matrices $(\s^i)_{AB}\equiv(\s^i\e)_{AB}$ obey
$(\s^i)_{AB}(\s^j)_{CD}=-i\e^{ijk}\e_{AC}(\s^k)_{BD}-{2\over
3}\d^{ij}\e_{AC}\e_{BD}$, where symmetrizations on $AB$ and $CD$ are
suppressed. The corresponding anti-hermitian generators $\hM_{ij}$ and
Dirac-matrices $\C^{ij}$ are related by $\C^{ij}\hM_{ij}=-iL_{AB}$, where
$L_{AB}\equiv 2i(\s^i)_{AB}L_i=(L^{AB})^\dagger$.} $\a=1,...,4$ and
$A=1,2$. We use the $SO(6)\sim SU(4)$ chiral notation, in which chiral
and anti-chiral spinors always have upper and lower indices,
respectively. No raising and lowering of chiral indices can be done,
while the doublet indices can of course be raised and lowered as usual.
Under hermitian conjugation, $(Q_\a^A)^\dagger=\bar Q^{\a A}$.
Multiplying \eq{q2} by $i\C^t$ from the right and projecting by $\Pi_+$
we find that the non-vanishing commutators are \bea
\{Q_\a^A,\bar Q^\b_B\} &=&2\d_\a^\b\d^A_BH' +{2\over
L}\,\d^A_B J_\a{}^\b -{i\over L}\,\d_\a^\b L^A{}_B\ ,\nn\\
{[}H',Q_{\a A}]&=& -{1\over 4 L}\, Q_{\a A}\ ,\quad\quad
{[}H',\bar Q^\a_A]= {1\over 4 L}\, \bar Q^\a_A\ ,\nn\\
{[}J_\a{}^\b ,Q_{\c A}]&=&\d^\b_\c\,Q_{\a A} -\frac{1}{4}
\d_\a^\b\,Q_{\c A} \ ,\qquad
{[}L_{AB},Q_{\c C}]=-2i\e_{C(A}Q_{\c B)}\ ,\label{susy2}\\
{[}J_\a{}^\b ,Q^\c_A]&=&\d^\c_\a\,Q^\b_A -\frac{1}{4}
\d_\a^\b\,Q^\c_A \ ,\qquad\quad
{[}L_{AB},Q^\c_C]=-2i\e_{C(A}Q^\c{}_{B)}\ ,\nn\\
{[}J_a{}^\b,J_\c{}^\d]&=&
\d_\a^\d\,J_\c{}^\b -\d_\c{}^\b\,J_\a^\d \ ,\nn\\
{[}L_{AB},L_{CD}]&=&-2i\e_{C(A}L_{B)D}+(C\leftrightarrow D)\ ,\nn
\eea
where $L_{AB}$ are the $SO(3)_\chi$ generators and we have defined the
$SO(6)$ generators $J_\a{}^\b \equiv -\frac{1}{4} M^{I'J'}
(\C_{I'J'})_\a{}^\b$. The non-trivial Jacobi identity $\{Q_{\a A},\{Q_{\b
B},Q^\c_C\}\}+{\rm cyclic}=0$ is indeed satisfied, and the fact that $H'$
does not commute with the supercharges is essential for this to happen.
Despite this non-commutativity the ground state energy may still vanish
\cite{baryon}.

For the two-spin solution, the superalgebra can be obtained using the
similarity transformation \eq{similarity}, noting that  $
{\mathbb S}^\dagger \C^{t}= \C^{t} {\mathbb S}^{-1}$.

\section{Vibration Spectrum}

In this section we examine the spectrum of small fluctuations around the
1/2 BPS single and two-spin solutions, including fermions for the $M2$ in
$AdS_7\times S^4$. For single-spin giants the scalar fluctuations were
analyzed in \cite{das} (the vibration spectrum for giants in the PP-wave
background was studied in \cite{sheik1,sheik2,sheik3}). There are several
motivations for this. Firstly, the solution defines the ground state for
a subsector of the $p$-brane field theory, based on normal coordinate
expansion, which is stable only if all fluctuation-modes have real and
positive frequencies. There is also the issue of singling out zero-modes
describing continuous shifts in the semi-classical parameters.

Secondly, the full normal-coordinate expansion, including loops,
becomes a $1/N$-expansion after absorbing a power of $\sqrt{N}$
into the fluctuation fields (the $\ell$-loop contributions to the
$n$-point diagrams scale like $N^{1-n/2-\ell}$), which reduces to
the free quadratic Lagrangian in the limit $N\to\infty$. Hence,
from the point of view of the dual CFT, the frequencies have a
direct interpretation in terms of the large $N$ limit of the
scaling dimensions of a subset of operators forming a ``tower'' on
top of a specific operator corresponding to the ground state of the
expansion on the $p$-brane \cite{BMN,berenstein,berkooz2}.
Another basic idea is that the tower has the structure
of a Fock space in the large $N$ limit. Indeed, as in the case of
the original giant gravitons \cite{das}, the vibration modes
of the two-spin giants have evenly spaced frequencies fixed by
the radii of curvature.

For the calculations in this section, we find it convenient to use
the following parametrization of the sphere $S^{m-2}$ in
(\ref{metrik1}): \be \label{spy}
d\O_{m-2}^2=(1-y^2)d\a^2+(\delta_{mn}+\frac{y_my_n}{1-y^2})dy^m
dy^n. \ee In these coordinates the $S^{m-2}$ part of the solution
\eq{sol} is given by \be \a=\tau/\tilde{L},\,\,\,\,\,y^{m}=0. \ee

\subsection{Bosonic Oscillations}

In the physical gauge that we employed, the coordinates can be
perturbed as
\bea
&&\phi=\tau/L+\delta\phi, \hs{4} \a=\tau/\tilde{L}+\delta\a, \nonumber\\
&&\theta=\theta_0+\delta\th, \hs{3} r=r_0+\delta r,
\hs{3} y^m=\delta y^m, \eea
where the fluctuations depend on all of
the world-volume coordinates. Expanding the action (\ref{act}) to
the linear order in perturbations we find that the variation
vanishes since the background obeys the field equations. On the
other hand, to the second order, the action becomes \be S_2=\int
d^{n-1}\sigma\,\sqrt{-\c^{(0)}}\,{\cal L}_2 \ee where \bea
&&2\,{\cal L}_2=\delta y^m
\left[\Box-\frac{1}{\tilde{L}^2\sin^2\th_0}\right] \delta y^m
+\delta r \left[\fr{\Box}{f_0}\right] \delta r + \delta\a\left[
(1+\fr{r_0^2}{\tilde{L}^2\sin^2\th_0})r_0^2\Box\right]\delta\a\nonumber\\
&&+\delta\phi\left[L^2\cot^2\th_0\Box\right]\delta\phi +\delta\th
\left[L^2\Box\right]\delta\th
+\frac{4r_0}{\tilde{L}\sin^2\th_0}\delta r\, \del_{\tau}\,\delta\a
\label{qa}\\
&&+\frac{2(n-3)L\cos\th_0}{\sin^3\th_0}\delta\th\,\del_\tau\,\delta\phi
+\frac{2(n-3)r_0^2\cos\th_0}{\tilde{L}\sin^3\th_0}\delta\th\,\del_\tau\,
\delta\a +\delta\a\left[\frac{2 r_0^2L}{\tilde{L}}\cot^2\th_0
\Box\right] \delta\phi,\nonumber \eea $\Box$ is the D'Alembertian
for the background world-volume metric $\c^{(0)}_{\a\b}$ which is
given by \be \c^{(0)}_{\a\b}=
\left( \begin{array}{cc} - \sin^2\th_0& 0 \\
0 & L^2\,\sin^2\th_0\,(g^\chi)_{ij}\end{array} \right) \ee and
$f_0=1+r_0^2/\tilde{L}^2$. In writing the above Lagrangian some
terms are integrated by parts; there is no surface contribution
coming from the spatial part of the world-volume since it is a
closed surface and the variations are assumed to vanish at
$\tau=\pm\infty$. We expand a generic perturbation as \be \delta X
= \sum_l \delta X_0 \,e^{i \o_l \tau}\,Y_l \ee where $Y_l$ are
spherical harmonics on the unit ($n-2$)-sphere obeying \be
(g^\chi)^{ij}\del_i\del_j\,Y_l=-Q_l\,Y_l, \ee with $Q_l=l(l+n-3)$.
Then, $\Box$ acting on the $l$'th mode becomes \be \Box\to
\fr{1}{\sin^2\th_0}(\o_l^2-\fr{Q_l}{L^2})\equiv D_l. \ee From the
above quadratic Lagrangian, we see that $\delta y^m$ perturbations
decouple and have the normal frequencies \be
\o_l^2=\frac{1}{\tilde{L}^2}+\frac{Q_l}{L^2}. \label{f1} \ee On
the other hand, $\delta r$, $\delta\a$, $\delta\phi$ and
$\delta\th$ modes are coupled. The resulting frequencies are
determined from the following equation \be
\left[\begin{array}{cccc}
\fr{D_l}{f_0} &  \frac{2i\o_lr_0}{\tilde{L}\sin^2\th_0} &0&0 \\
-\frac{2i\o_lr_0}{\tilde{L}\sin^2\th_0} &
(1+\fr{r_0^2}{\tilde{L}^2 \sin^2\th_0})r_0^2 D_l
&\frac{r_0^2L}{\tilde{L}} \cot^2\th_0 D_l
&-\fr{i\o_l(n-3)r_0^2\cos\th_0}{\tilde{L}\sin^3\th_0}\\
0 &\frac{r_0^2L}{\tilde{L}}\cot^2\th_0 D_l & L^2 \cot^2\th_0 D_l
&-\fr{i\o_l(n-3)L\cos\th_0}{\sin^3\th_0}\\
0&\fr{i\o_l(n-3)r_0^2\cos\th_0}{\tilde{L}\sin^3\th_0}
&\fr{i\o_l(n-3)L\cos\th_0}{\sin^3\th_0} & L^2 D_l
\end{array} \right]
\left[\begin{array}{cccc} \delta r \vs{.1} \\ \delta\a \vs{.1}\\
    \delta\f \vs{.1} \\ \delta
\th \end{array} \right]=0. \label{m} \ee Calculating the
determinant, we see that (for non-zero $r_0$, $\cos\th_0$ and
$\sin\th_0$) it factorizes into two quadratic equations for
$\o_l^2$ from which the  following normal frequencies can be
obtained \bea \o^2_{l\pm}&=&\frac{1}{L^2}\left[Q_l+
\frac{(n-3)^2}{2}\pm
  (n-3)\sqrt{Q_l+\frac{(n-3)^2}{4}}\,\,\right]\label{f2}\\
 \o^2_{l\pm}&=&\frac{1}{\tilde{L}^2}\left[\frac{\tilde{L}^2}{L^2}Q_l+ 2 \pm
  2 \sqrt{\frac{\tilde{L}^2}{L^2}Q_l+1}\,\,\right].\label{f3}
\eea There is no unstable mode in the system since all $\o_l^2$
are real and non-negative. Using the fact $Q_l=l(l+n-3)$,
(\ref{f2}) simplifies to $\o_{l+}=(l+n-3)/L$ and $\o_{l-}=l/L$.

Frequencies (\ref{f1}) and (\ref{f2}) constitute the vibration
spectrum of the single-spin giant wrapped in $S^n$
\cite{mathur}. Therefore, the modes (\ref{f3}) can be thought to
arise due to the spin in $AdS$. The rotation disappears when
$r_0\to0$ and the single spin solution is recovered. In this limit
($r$, $\a$, $y^m$)  coordinate system is not well defined. To read
the eigenfrequencies one should either introduce flat coordinates
or just treat $(r_0\delta\a)$ as the true perturbation. In the
later case, it is easy to see that $\delta r$ and $r_0\delta \a$
perturbations decouple from $\delta\phi$ and $\delta\th$ modes and
(\ref{f3}) should be replaced with (\ref{f1}) \footnote{Another
limit that one may wish to consider is
  when the rotation in $\phi$
  disappears, i.e.  $\th_0\to\pi/2$. In this case, $P_\phi$ does not
  vanish, on the contrary it reaches its maximum \eq{mphi}.
  This is the maximal
  giant and the angular momentum arises due to the coupling of the
  background flux to the brane. The spectrum does not change. On the
  other hand in $\th_0\to0$ limit the brane collapses to a point and
  fluctuation analysis becomes ill defined. Indeed, before reaching
  this value, the world-volume becomes highly curved and the probe
  brane approximation fails.}.

It is remarkable that for $(m,n)=\{(5,5),(4,7),(7,4)\}$,
$L=(n-3)\tilde{L}/2$ and (\ref{f2}) becomes equal to (\ref{f3}).
In this case, the eigenfrequencies are given by \be \label{fb}
\o_l=\fr{l+(n-3)/2}{L},\hs{5}\o_{l+}=\fr{(l+n-3)}{L},\hs{5}\o_{l-}
=\frac{l}{L}, \ee where the first $\o_l$ is for $\delta y^m$ with
degeneracy $(m-3)$ and the others are for the mixing of $\delta
r$, $\delta \a$, $\delta\f$, $\delta \th$, each frequency
occurring with degeneracy 2. The eigenfrequencies for  the single
spin giant is given by (\ref{fb}) but the degeneracies are
$(m-1)$, 1, 1, respectively.

\subsection{Bosonic Zero Modes and Spectrum}

As discussed in \cite{mathur}, some of the above excitations (zero
modes) correspond to the collective motion of the brane since
there are continuous families of equilibrium configurations. These
modes will change the quantum numbers (i.e. the conserved
quantities like $P_\phi$ and $P_\a$) and should be removed from
the spectrum since they can no longer be viewed to belong to the
giant we started with. To linear order, there can be no change in
the conserved quantities due to $l\not=0$ modes, since these are
calculated at a fixed world-volume time as integrations over the
sphere and we have $\int Y_l=0$ when $l\not=0$. Note that $Y_0$ is
the constant harmonic and we have $Q_0=0$.

For $l=0$, we see from (\ref{f1}) that the frequencies for $\delta
y^m$ perturbations become $\o_0=1/\tilde{L}$. These modes
correspond to the shifts in the great circle in $S^{m-2}$. To see
this, following \cite{mathur}, one can embed $S^{m-2}$ in
$(m-1)$-dimensional flat space with coordinates $x_1,x_2,y^m$. The
unit sphere can be defined as $x_1^2+x_2^2+y^my^m=1$. The
coordinate system (\ref{spy}) corresponds to the parametrization
\be
x_1=\cos\b\cos\a,\,\,\,\,x_2=\cos\b\sin\a,\,\,\,\,y^my^m=\sin^2\b
\ee The brane is circling in ($x_1$-$x_2$) plane at $\b=0$ (and
thus $x_1=\cos\a$, $x_2=\sin\a$ and $y^m=0$). One can now rotate
this plane by an angle $\delta$: \be
x_1'=x_1\cos\delta-y_1\sin\delta,\,\,\,\,\,\,\,\,
y_1'=x_1\sin\delta+y_1\cos\delta. \ee Recalling that $x_1=\cos\a$
and $y^m=0$, for small $\delta$ one has \be
x_1'=x_1,\,\,\,\,\,\,\,\,y_1'=\delta \cos(\tau/\tilde{L}) \ee This
change in $y_1$ is precisely equal to the perturbation with $l=0$.
Thus these modes represent shifts in the direction of $P_\a$
without changing its magnitude, and $l=0$ frequencies in
(\ref{f1}) should be removed from the spectrum.

From (\ref{f2}) and (\ref{f3}), there are four more frequencies
with $l=0$. These are $\o_0^{(1)}=\o_0^{(2)}=0$,
$\o_0^{(3)}=(n-3)/L$  and
  $\o_0^{(4)}= 2/\tilde{L}$. The modes corresponding to $\o_0^{(1)}$
and $\o_0^{(2)}$ represent shifts in the values of $r_0$ and
$\th_0$. From (\ref{mphi}) and (\ref{ma}), a change in $r_0$
modifies $P_\a$ and a change in $\th_0$ alters both $P_\phi$ and
$P_\a$. Therefore, these modes should not be in the spectrum.

On the other hand, the perturbations corresponding to $\o_0^{(3)}$
and $\o_0^{(4)}$ respectively obey \bea &&\delta
r=\delta\a=0,\hs{5}\delta\phi=i(\tan\th_0)\,\delta\th,\label{z1}\\
&&\delta
r=\frac{iL(\tilde{L}^2+r_0^2)}{\tilde{L}r_0}\,\delta\phi,\hs{5}
\delta\a=-\frac{L\tilde{L}}{r_0^2}\,\delta\phi,\hs{5}\delta\th=0.\label{z2}
\eea It is now straightforward to verify that the change in
$P_\phi$ and $P_\a$ is zero under these perturbations. For
example, from (\ref{m1}) the variation of $P_\phi$  under
(\ref{z1}) becomes \be\label{z3} \delta P_\phi=\frac{\del
P_\phi}{\del \th}\,\delta\th+\frac{\del
  P_\phi}{\del\dot{\phi}}\,\delta\dot{\phi}.
\ee Using $\delta\dot{\phi}=i\o_0^{(3)}\delta\phi$ and (\ref{z1})
one finds $\delta P_\phi=0$. Therefore, the zero modes
corresponding to $\o_0^{(3)}$ and $\o_0^{(4)}$ should be kept in
the spectrum.

After the elimination of these zero modes, we end up with the
following spectrum of small scalar field fluctuations for
$(m,n)=\{(5,5),(4,7),(7,4)\}$ given in units of $1/L$ (note that
$l$ is shifted compared to \eq{fb}): \vs{3}
%\begin{table}\label{t1}
\begin{center}
\begin{tabular}{|c|c|c|c|}
\hline
%&&&\\
Bosonic $\o_l$ ($l\geq 0$)  &\hs{3}$l+(n-1)/2$ \hs{3}&\hs{3}
$l+n-3$ \hs{3} &\hs{3} $l+1$ \hs{3}\\
%&&&\\
\hline
%&&&\\
Multiplicity (two-spin) & $(m-3)_{l+1}$ & $(2)_l$ & $(2)_{l+1}$\\
%&&&\\
\hline
%&&&\\
Multiplicity (single-spin) & $(m-1)_{l+1}$ & $(1)_l$ & $(1)_{l+1}$\\
%&&&\\
\hline
\end{tabular}
\end{center}
%\end{table}
%
where the suffix indicates the leading $SO(n-1)_\chi$ highest
weight label.

\subsection{Fermionic Oscillations of $M2$-Brane (Two-spin)}

We calculate the spectrum of the fermionic oscillations for an
$M2$ brane in $AdS_7\times S^4$. For the membrane, the quadratic
action for the fermion fluctuations in a general bosonic
background of $D=11$ supergravity was derived in \cite{berg} which
can be written as \be\label{lt} {\cal L}_{\Th}=i
\sqrt{-\gamma}\,\c^{\a\b}\,\del_{\a}\,X^M \,\,E_M^A\, \bar{\Th}
\,(I-\C)\,\C_A\,\tilde{\nabla}_{\b}\Th \ee where \be\label{53}
\tilde{\nabla}_{\a}=\del_{\a}\,+\,\del_{\a}X^M\left[\fr{1}{4}\o^{AB}{}_{M}
\C_{AB}
+\fr{1}{288}(8\delta_{M}^{P}\C^{QRS}-\C_{M}{}^{PQRS})H_{PQRS}\right],
\ee $\c_{\a\b}$ is the induced metric and $E^A_M$ is an
orthonormal basis in the space-time. The non-vanishing components
of the connection one-forms\footnote{Our convention is
$de^A+\o^A{}_B\wedge e^B=0$ where we expand
$\o^A{}_B=\o^A{}_{BC}\,e^C$.} that contribute to (\ref{53}) are
\bea \label{nzcon}
&&\o^{\hat{\f}}{}_{\hat{\th}\hat{\f}}=-\fr{\tan\th_0}{L},\hs{5}
\o^{\hat{i}}{}_{\hat{\th}\hat{j}}=-\fr{\cot\th_0}{L}\delta^{i}{}_{j},\hs{5}
\o^{\hat{i}}{}_{\hat{j}\hat{k}}=
\fr{\o^{(1)\hat{i}}{}_{\hat{j}\hat{k}}}{L\sin\th_0},\\
&&\o^{\hat{t}}{}_{\hat{r}\hat{t}}=\fr{r_0}{\tilde{L}^2\sqrt{f_0}},\hs{5}
\o^{\hat{\a}}{}_{\hat{r}\hat{\a}}=\fr{\sqrt{f_0}}{r_0},\nn \eea
where a subscript '(1)' on a quantity indicates that it is defined
on a unit 2-sphere. Using the solution (\ref{sol}) and
$H_{\hat{\th}\hat{\f}\hat{1}\hat{2}}=3/L$ we find \bea
\tilde{\nabla}_\tau&=&\del_\tau\,+\, \fr{1}{2\tilde{L}}\left[
\fr{r_0}{\tilde{L}}\C_{t} + \sqrt{f_0} \C_\a\right]\C^r -
\fr{1}{4L}\left[ \fr{r_0}{\tilde{L}}\C_{\a} + \sqrt{f_0}
  \C_t \right]\c\nn\\
&-&\fr{1}{2L}\left[\sin\th_0\,\C^{\f}\,+\,
\cos\th_0\C^{12}\right]\C^{\th} \label{62}\\
\tilde{\nabla}_i&=&\nabla^{(1)}_{i}\,-\fr{1}{2}\left[\cos\th_0 \,
\C^{\th}\,+ \sin\th_0\,\c\right]\C^{(1)}_{i}    \label{63}\\
\C&=&\fr{1}{\sin\th_0}\left[\sqrt{f_0}\,\C_{t12}\,+\,\fr{r_0}{\tilde{L}}
\,\C_{\a 12}\,+\, \cos\th_0\, \C_{\f12}\right]\label{57} \eea
where $\c=\C_{\th\f12}$, indices (1,2) refer to $\chi_1$ and
$\chi_2$ directions and recall that the indices on the gamma
matrices are flat. To fix $\kappa$ symmetry, we impose
$\C\Th=-\Th$ which also gives $\bar{\Th}\C=-\bar{\Th}$. Using this
gauge condition in (\ref{lt}) we obtain \be\label{ara1} {\cal
L}_{\Th}=-i L^2 \sin^2\th_0 \sqrt{-\c^{(1)}}\left[
\bar{\Th}\C_{12}\tilde{\nabla}_{\tau}\Th -
\fr{1}{L}\bar{\Th}\C_{(1)}^{i}\tilde{\nabla}_i\Th\right]. \ee To
proceed we use the following representation of the 11-dimensional
gamma matrices: \bea &&\C_t=i\s_2\ot \c_5 \ot \s_3\ot\s_3, \hs{5}
\C_r=\s_3\ot\c_5 \ot
\s_3\ot\s_3, \nn\\
&&\C_\a=\s_1\ot \c_5 \ot \s_3\ot\s_3, \hs{5}\, \C_m=I_2\ot\c_m\ot
\s_3\ot\s_3,\label{gr}\\
&&\C_\th=I_2\ot I_4 \ot \s_2\ot\s_3, \hs{5} \,\,\,\C_\f=I_2\ot I_4
\ot
\s_1\ot\s_3,\nn\\
&&\C_i=I_2\ot I_4 \ot I_2\ot\s_i,\hs{8} C=i\sigma^2\ot c \ot
\sigma^1 \ot i\sigma^2,\nn \eea where $i=1,2$,\, $m=1,..,4$ and
$\gamma^m c$ are symmetric. Then (\ref{57}) become \be
\C=\fr{i}{\sin\th_0}\left[(\fr{r_0}{\tilde{L}}\s_1+i\sqrt{f_0}\,
  \s_2)\ot\c_5\ot\s_3\ot I_2\,+\,\cos\th_0 \,\,
 I_2\ot I_4\ot\s_1\ot I_2 \right]
\ee and writing ${\cal L}=2{\bar\Th}M\Th$ we find \bea M&=&I_2\ot
I_4\ot I_2\ot \left(\s_3\del_\tau+
\fr{i}{L}\s^i\nabla_i^{(1)}\right)\,
-\,\fr{1}{4L}(\fr{r_0}{\tilde{L}}\s_1+i\sqrt{f_0}\,
  \s_2) \,\ot (I_4+\c_5)\ot I_2\ot \s_3 \nn\\
&+&\fr{i\sin\th_0}{2L}\,I_2\ot I_4\ot \s_3 \ot \s_3
+\fr{i\cos\th_0}{2L}\,I_2\ot I_4\ot \s_2 \ot \s_3. \label{mass1/4}
\eea
We solve the $\C$ projection condition on $\Th$ as
\bea \Th
&=& \sum_{\e,\e'=\pm 1} (P^{(1)}_{\e}\ot
P^{(2)}_{\e'}\ot P^{(3)}_{\e\e'}\ot I_2)\Th_{\e,\e'}\ ,\label{th1}\\
P^{(1)}_{\e}&=&{1 \over 2}(I_2+\e(i{r_0\over
\tilde{L}}\s_1-\sqrt{f_0}\s_2))\,
\label{th2}\\
P^{(2)}_{\e'}&=&{1 \over 2}(I_4+\e'\c^5)\ ,\label{th3}\\
P^{(3)}_{\e\e'}&=& {1\over 2}(I_2
-\e\e'(\s_3+i\cos\th_0\s^1)/\sin\th_0) \label{th4} \eea where
$\Th_{\e,\e'}$ are unrestricted. Each term in the sum is a product
of three projectors of half maximal rank, which implies that $28$
of the $32$ components in each $\Th_{\e,\e'}$ are set to zero,
such that $\Th$ has $4\times 4=16$ components. From \eq{th2} and
\eq{th4} one verifies \bea ({r_0\over
\tilde{L}}\s_1+i\sqrt{f_0}\s_2)P^{(1)}_{\e} &=& -i\e
P^{(1)}_{\e}\ ,\\
P^{(3)}_{\e\e'}(\cos\th_0\s_2+\sin\th_0\s_3)P^{(3)}_{\e\e'} &=&
-\e\e' P^{(3)}_{\e\e'}, \eea which shows that the masses are
independent of both $r_0$ and $\th_0$, while of course the
direction in spinor space of the projected $\Th$ depends on $r_0$
and $\th_0$ via \eq{th2} and \eq{th4}. To extract the normal
frequencies we expand  $\Th_{\e,\e'}$ in terms of spherical spinor
harmonics  on $S^2$ and substitute
\be \del_\tau\to i \o_l \hs{10} i\s^i\nabla_i^{(1)}\to
(l+1/2)\s_1,\hs{5}l=1/2,3/2,5/2,... \ee
The resulting characteristic equation has the matrix
\bea M&=& iP^{(1)}_\e\ot P^{(2)}_{\e'} \ot P^{(3)}_{\e\e'}\ot
\left((\o_l +{1\over 4L}\e(1-\e'))\s_3 -{i\over
L}(l+1/2)\s_1\right) \eea
which yields the (positive\footnote{There are also conjugate modes
  which have negative frequencies with the same degeneracy})
eigenfrequencies ($l=1/2,3/2,5/2,\dots$)
\be\label{ff}
\o_l=\fr{(l+1/2)}{L}\,\,\{\e'=1\},\hs{3}\o_{l}=\fr{(l+1)}{L}\,\,\{\e'=\e=-1\},
\hs{3}\o_{l}=\frac{l}{L}\,\,\{\e'=-1,\e=1\}. \ee
The equations of motion (note that the Dirac operator projects by half in
the last slot) eliminate half of the fermionic degrees of freedom, which
means that the first $\o_l$ occurs with degeneracy 4 and the others with
degeneracy 2, giving 8 frequencies for each $l$, forming four doublets
$(2_{\e\e'})_l$ under the $SO(3)_{\e'}\subset SO(4)$, where $SO(4)$ is
the manifest part of the unbroken $SO(6)$.

\subsection{Fermionic Oscillations of $M2$-Brane (Single-spin)}

For the 1/2 supersymmetric single-spin giant gravitons the
spectrum of the fermionic oscillations has not been examined
before. In this subsection we fill this gap for a giant $M2$ in
$AdS_7\times S^4$. The calculation is very similar to the two-spin
membrane studied above. Mainly one should take
$r_0\to0$ limit and remember the fact that $\a$ is now equal to a
constant which makes a difference when pulling back the objects to
the world-volume. We find that only the first line of \eq{nzcon}
contributes to \eq{53}. Also, \eq{62} becomes \be
\tilde{\nabla}_\tau= \del_\tau\,\,- \fr{1}{4L}\,
  \C_t \c\,-\,\fr{1}{2L}\left[\sin\th_0\,\C^{\f}\,+\,
\cos\th_0\C^{12}\right]\C^{\th}, \ee and \eq{63} is not modified.
Moreover, in \eq{57} one should set $r_0=0$ which shows that $\C$
projection condition on $\Th$ can be solved as above. However, the
mass matrix \eq{mass1/4} changes to \bea M&=&I_2\ot I_4\ot I_2\ot
\left(\s_3\del_\tau+ \fr{i}{L}\s^i\nabla_i^{(1)}\right)\,
-\,\fr{i}{4L}\,\s_2\,\ot\,\c_5\,\ot I_2\ot \s_3 \nn\\
&+&\fr{i\sin\th_0}{2L}\,I_2\ot I_4\ot \s_3 \ot \s_3
+\fr{i\cos\th_0}{2L}\,I_2\ot I_4\ot \s_2 \ot \s_3. \label{mass1/2}
\eea
After expanding $\Th$ in terms of spherical spinor harmonics on
$S^2$ we obtain the eigenfrequencies ($l=1/2,3/2,\dots$)
\be\label{ffs} \o_{l}=\fr{(l+3/4)}{L}\,\,\{\e'\times\e=1\},
\hs{3}\o_{l}=\frac{(l+1/4)}{L}\,\,\{\e'\times\e=-1\}, \ee
where each $\o_l$ occurs on-shell with degeneracy 4, giving rise
to two $SO(6)\simeq SU(4)$ spinors, that we shall denote by $(\bar
4)_l$ ($\e\e'=1$) and $(4)_l$ ($\e\e'=-1$).

\subsection{Fermionic Zero Modes and Spectrum}

As for bosons,  some of these fermionic oscillations  correspond
to the collective motion of the giant graviton (in spinor space)
and should be removed from the spectrum. These are precisely the
modes generated by the broken supersymmetries, i.e. $\Th=(1-\C)\e$
where $\e$ is the space-time Killing spinor $\e(X)$ in \eq{kilsip}
evaluated on the membrane. Let us emphasize that such a mode does
not necessarily obey the equations of motion (the spinor $\e$
satisfies $\tilde{\nabla}_\b\e=0$, however one may have
$[\C^{\b}\tilde{\nabla}_\b,\C]\not=0$), so one should directly
examine the field equations to extract zero modes.

In our case, there is a short way to proceed; for a maximal giant
$(\th_0=\pi/2)$  we have $[M,\C]=0$ where $M$ is the operator
given in \eq{mass1/2} and \eq{mass1/4}. Therefore
$M(1-\C)\e=(1-\C)M\e=0$, so the zero modes are given by the
spinors $\e$ obeying $(1+\C)\e=0$.

For the single-spin maximal giant, it is easy to show that
$(1\pm\C)\e=0$ for $(1\mp\C^{t\f})\e_0=0$ where $\e_0$ is the
constant spinor in \eq{kilsip}. So, all 16 modes generated by the
broken supersymmetries $\C^{t\f}\e_0=\e_0$ are zero modes. Half
of them obeying $\c\C^\f \e_0=\e_0$ have the frequency $3/(4L)$
and other half have $-3/(4L)$. Comparing with \eq{ffs}, one finds
that the $l=1/2$ modes with $\e'\e=-1$ should be eliminated.

For the two-spin maximal giant, the projection $(1+\C)\e=0$ can be
solved as in
subsection \ref{susysection}. With the notation used in that subsection,
one finds that the zero modes are generated by the spinors $\e^{+-}$
and $\e^{++}$ (here $\pm$ assignments on $\e_0$ are different than
$\e$ and  $\e'$ values). Decomposing further, the fermions generated by
$\c\C^\f\,\e^{+-}\,=\,\pm\,\e^{+-}$ give the  zero modes with the
frequencies $\omega=\pm 1/(2L)$. On the other hand, the fermions generated by
$\C^{\a_4\a_5}\,\e^{++}\,=\,\pm\,\e^{++}$ give the  zero modes with
$\omega= \pm 1/L$.  From \eq{ff}, we see that the $l=1/2$ modes
with $\e=1,\e'=-1$ and the modes with $\e=-1$,$\e'=1$ should be
removed from the spectrum.\footnote{As shown above, for the single-spin
  solution the zero modes have the chiralities $\e\e'=-1$. As
  $r_0\to0$ the two-spin solution reduces to the single-spin one
  and thus its zero modes should have the same chiralities
  $\e\e'=-1$.}
Summarizing, we have the following eigenfrequencies after the
elimination of the zero modes (in units of $1/L$ and where  the
suffices indicate $SO(3)_\chi$ spin and $(\e,\e')$ assignments):
\vs{3}
%\begin{table}\label{t1}
\begin{center}
\begin{tabular}{|c|c|c|c|}
\hline Fermionic $\o_l$ ($l\geq 1/2$)
&$l+1+\eta/4$ & \hs{3}$l+1/2+\eta/4$ \hs{3}
& \hs{3}$l+3/2$ \hs{3} \\
\hline Multiplicity (two-spin) $\eta=0$ & $(2_{\scriptscriptstyle
+-})_{l+1} +(2_{\scriptscriptstyle --})_{l}$ &
$(2_{\scriptscriptstyle ++})_l$ & $ (2_{\scriptscriptstyle -+})_{l+1}$ \\
\hline Multiplicity (single-spin) $\eta=1$ & $(4)_{l+1}$ & $(\bar
4)_l$ & \\
 \hline
\end{tabular}
\end{center}
%\end{table}

\subsection{Comments on the Supermultiplet Structure}

The vibration spectra of single and two-spin 1/2 BPS membranes in
$AdS_7\times S^4$ can  be arranged into multiplets  of the unbroken
supersymmetry algebras found in subsection \ref{alg}. Here, we would
like to comment on some salient features of this computation and give
the multiplet for the single-spin solution.

Let us start with the bosonic fluctuations. For the two-spin solution,
$(\d r,\d\alpha,\d \theta,\d \phi)$-sector of the bosonic quadratic
Lagrangian can be diagonalized by introducing two complex fields $\d
z(\tau,\sigma^i)$ and $\d w(\tau,\sigma^i)$. These fields must be
complex due to the first order time-derivatives, and can be chosen as
\be \d z=L(\d\theta+i\d\phi/\cos\th_0)+{\cal O}(\cos\th_0)\
,\qquad \d w=\d r+i\tilde L\d\alpha+{\cal O}(\cos\th_0)\ . \ee
Since the frequencies are independent of $\th_0$, it suffices to
consider the limit $\sin\th_0\rightarrow 1$ (keeping $\d z$
fixed). Performing the expansion using the harmonics $Y_l$ on $S^2$
gives the frequencies as:
\bea z_l(\tau)&=&e^{i\omega_l^+\tau}
a_l^\dagger+e^{-i\omega_l^-\tau} b_l\ ,\label{freqshiftz}\\
w_l(\tau)&=&e^{i\omega_l^+\tau} c_l^\dagger+e^{-i\omega_l^-\tau}
d_l\ , \eea
where $\o_l^-=l/L$ and $\o_l^+=(l+1)/L$. Note the shift in the
negative frequency relative to the positive frequency. The
conjugate oscillators are contained in $(\d
z(\tau,\sigma^i))^\dagger$ and $(\d w(\tau,\sigma))^\dagger$,
which have the frequency parts
\bea \bar z_l(\tau)&=&e^{i\omega_l^-\tau}
b_l^\dagger+e^{-i\omega_l^+\tau} a_l\ ,\label{freqshiftz2}\\
\bar w_l(\tau)&=&e^{i\omega_l^-\tau}
d_l^\dagger+e^{-i\omega_l^+\tau} c_l\ . \eea
There is no corresponding shift in the real $\d y^m$-fields, which
transform as a $4$-plet under $SO(4)$. For the single-spin solution,
$\d r$ and $\d \a$ perturbations combine with $\d y^m$-fields to form
a $6$-plet under $SO(6)$. Here, there is no need to introduce the
complex scalar $\d\omega$ and the expansion of $\d z$ is identical to
\eq{freqshiftz}.

In the fermionic sector, firstly eq. \eq{M2fermicharge} implies
that $U(1)_S\times SO(4)\times U(1)_J\times SO(3)_\chi$ (two-spin) or
$SO(6)\times U(1)_J\times SO(3)_\chi$ (single-spin) rotations
of the unbroken supercharges $Q$ are related by conjugation by
$g(X_0)$ to rotations of $\Theta$. The former representation is
generated by the set of Dirac matrices found in the anti-commutator
(see e.g. \eq{qq1/2}), while the latter representation is diagonal
with respect to the decomposition defined by $P^{(1)}_\e\otimes
P^{(2)}_{\e'}\otimes P^{(3)}_{\e\e'}\otimes I_2$ introduced in Section
3.3 (such that $U(1)_S$ acts in the first slot or $SO(6)$ acts in the
first two slots and so on in the indicated order). Hence, for the
two-spin solution using $SO(4)=SO(3)_+\times SO(3)_-$ and the notation
where $(\a,\dot a)$ are the doublet indices of $SO(3)_+\times
SO(3)_\chi$, the fermionic $(2_{\pm+})_l$ states are contained in
the mode-expansion of a complex fermion $\Th_{\a}(\tau,\sigma^i)$
and its hermitian conjugate $\bar{\Th}_\a(\tau,\sigma^i)$, while the
$(2_{\pm-})_l$ states are contained in $\Th_{\dot\a}(\tau,\sigma^i)$ and
$\bar\Th_{\dot\a}(\tau,\sigma^i)$. Similarly, for the single-spin
solution $(4)_{l+1}$ states are contained in the mode expansion of a
complex $SU(4)$ 4-plet $\Theta_\a$ and $(\bar{4})_l$
states are contained in $\bar 4$-plet $\bar \Theta^\a$.

Next, we turn to the identification of the $S$ and $J$ charges.
From the geometric sigma-model picture, it follows that a full
$U(1)_J$-transformation generated by
$J=J_{(0)}+J_{(1)}+J_{(2)}+\cdots$, decomposes into a constant shift
of $\phi$ followed by a rotation (by the same angle) in the tangent
space spanned by $(\d\theta,\d\phi)$. In the linearized theory the
shift is generated by the zero-mode in $J_{(0)}$ while the tangent
space rotation is generated by $J_{(2)}$. Similarly, $S_{(1)}$ shifts
$\alpha$, and $S_{(2)}$ rotates $(\d r,\d\alpha)$. Hence $J_{(2)}$ and
$S_{(2)}$ generate symmetries of the quadratic action. Clearly, these
charges can be computed from the normal coordinate expansion, though
this is an unnecessarily tedious procedure, given the fact that the
charges of all fluctuations are fixed (up to an overall sign) by
supersymmetry.

Let us illustrate this for the single-spin solution.
Recall that the spectrum now consists of a real $SO(6)$-vector $\d y^{I'}$
($I'=1,\dots,6$), a complex scalar $\d z$ ($J=1$), a $4$-plet
$\Theta_\a$ ($J=-1/2$) and a $\bar 4$-plet $\bar \Theta^\a$
($J=1/2$). As we will show, the supersymmetry is consistent with the
charge assignments. Note that the positive frequencies in $\d z$ and
$\d\bar z$ are given by $\o_l^+=(l+1)/L$ and $\o_l^-=l/L$,
respectively, and that the negative ones are shifted. In
\eq{susychargesingle} we found unbroken supercharges $Q_\a^A$
($(E,J)=(-\fr{1}{4L},\fr{1}{2})$) and $\bar Q^\a_A$
($(E,J)=(\fr{1}{4L},-\fr{1}{2})$) transforming as $(4,2)$ and
$(\bar 4,2)$ under $SO(6)\times SO(3)_\chi$. The spectrum of
non-zero modes now fits into a single tower ($l=0,1,2,\dots$):
\bea &&(1_1;(l+1)_l)\
\longrightarrow^{\!\!\!\!\!\!\!\!{Q^\dagger}}\ (\bar
4_{1/2};(l+5/4)_{l+1/2})\
\longrightarrow^{\!\!\!\!\!\!\!\!{Q^\dagger}}\
(6_0;(1+3/2)_{l+1})\nn\\&&\longrightarrow^{\!\!\!\!\!\!\!\!{Q^\dagger}}
\ (4_{-1/2};(l+7/4)_{l+3/2})\
\longrightarrow^{\!\!\!\!\!\!\!\!{Q^\dagger}} \ (
1_{-1};(1+2)_{l+2}),\eea
where the quantum numbers are listed as $(R_J;(L\o)_l)$, where $R$
is $SO(6)$ irreps.

There remains the following zero-mode oscillators: \be b^\dagger
(\o=0)\ ,\qquad y^\dagger_{I'}(\o=1/2)\ ,\qquad \Th^{\dagger \a}_A
(\o=3/4)\ ,\qquad b^\dagger_{AB} (\o=1)\ .\ee
Since $b^\dagger$ has vanishing frequency and $SO(6)\times
SO(3)_\chi$ charges, it is a supersymmetry singlet. The remaining
states form a multiplet (with $9$ bosons and $8$ fermions) with
supercharge and bosonic generators given by \bea Q_{\a
A}&=&{1\over \sqrt L}\left(2b^\dagger_A{}^B\Th_{\a
C}+(\c^{I'})_{\a\b}y_{I'}\bar\Th^{\dagger \b}_A\right)\ ,\nn\\
H'&=&{1\over L}\left(b^{\dagger AB}b_{AB}+{3\over
4}\bar\Th^{\dagger\a}_A\Th_\a^A+{1\over 2}y^\dagger_{I'}
y_{I'}\right)\ ,\\
\hM_{I'J'}&=& -\left(2y^\dagger_{[I'}y_{J']}+{1\over
2}\bar\Th^{\dagger \a}_A
(\c_{I'J'})_\a{}^\b\Th_\b^A\right)\ ,\nn\\
L_{AB}&=&2i\left(2b^{\dagger}_{(A}{}^C b_{B)C}+\bar\Th^{\dagger
\a}_{(A}\Th_{B)\a}\right).\nn\eea For the two-spin solution a similar
analysis can be repeated, and it would be interesting to examine how the
similarity transformation \eq{similarity} connects the two multiplet
structures.

\section{General Spherical Giants}

In this section we examine the most general ansatz for a spherical
giant configuration. After identifying the brane directions
in space-time, the spherical symmetry implies that all coordinate fields depend
only on time. Hence, while their shape is fixed, these branes can
``breathe'' and perform point-like motion in the remaining
directions.  As we shall  see, the whole system consists of a warped
product of a breathing mode and a relativistic point particle on $S^n$
(for the branes expanding in $AdS$) or  in $AdS_m\times S^1$ (for  the
branes expanding in the sphere)  with an additional
velocity dependent potential on $S^1$. The warping means that the
point particle motion decouples from the breathing in a suitable
world-volume time.  We show that the equations of motion  are
integrable. This can easily be verified  using  flat embedding
coordinates  where the  solution takes  a simple
form.   In terms  of the  usual  spherical or  $AdS$ coordinates,  one
obtains first  order equations that are  nested such that  they can be
integrated  further  one by  one.   Moreover,  switching to  canonical
fields leads to complete  separation of variables, with the
emergence  of relatively simple  potentials, known  as P\"oschl-Teller
Type I and II, for all point-particle coordinates, while the breathing
mode is governed by a seemingly more complicated potential.

We also derive BPS bounds for the energy as a function of
the constants of motion, and show that these are saturated only by
the 1/2 supersymmetric single-spin and the two-spin giants found in
section 2. The quantization of these objects and some comparisons with
the CFT side are  discussed in subsections \ref{quantizationsph} and
\ref{hol}.

To facilitate the analysis we use the $AdS$ metric
\be ds^2= \tilde{L}^2\, \left(-\cosh^2r\, dt^2 + dr^2 + \sinh^2
r\, d\O_{m-2}^2 \right) + L^2 d\O_n^2\, , \label{metric}\ee
where the line-elements on the unit spheres are specified in more
detail below. The $t$ coordinate is now dimensionless, which brings an
extra $\tilde{L}$ factor in time-derivatives when compared to some of
our previous results.

\subsection{Branes Expanding in $AdS$ (Electric)}

\label{giantads} In this case the background supports an $m$-form
field strength which is given in the tangent space basis by
$H_{\hat{r}\hat{t}\hat{\a}_1..\hat{\a}_{m-2}}=(m-1)/\tilde{L}$
(see also footnote \ref{ft}). We choose the static gauge and
identify the $(m-2)$-brane world-volume coordinates $\s_i$ with
the coordinates of $S^{m-2}$ in \eq{metric}:
\be t=\s_0 = \t, \hs{5} \a_1=\s_1,\,\,...\,\,, \a_{m-2}=\s_{m-2}\, .
\ee
We write the metric on the unit $n$ sphere in \eq{metric} as
\be\label{4sph}
d\O_n^2=G_{ab}\,d\f^a\,d\f^b,\hs{7}a,b=1,..,n,
\ee
and assume a solution of the form:
\be\label{an1}
r=r(\t), \hs{5}  \f^a=\f^a(\t).
\ee
This is a generalization of the dual giant configurations found in
\cite{myers,hashimoto}. The pull-back of the spacetime metric
to the  world-volume is given by:
\be \g_{\a\b}=\partial_{\a}X^M\partial_{\b}X^N g_{MN}=
\left( \begin{array}{cc} - \D^2 & 0 \\
0 & \tilde{L}^2\,{\rm sinh}^2r \ (g^\a)_{mn}\end{array} \right)
\ee
where $(g^\a)_{mn}$ is the metric on the unit $(m-2)$-sphere and
\be \D^2= \tilde{L}^2\,({\rm cosh}^2{r} - \dot{r}^2)-
L^2\,G_{ab}\dot{\f}^a\dot{\f}^b.
\label{delta}
\ee
In this section, dot always denotes derivative with respect to $\t$. In
(\ref{field}) for $M=\a_m$ the equations are obeyed trivially and for
$M=t$ one gets
\be
\D=\frac{\tilde{L}\, ({\sinh r})^{m-2} \ \cosh^2r}{({\sinh
r})^{m-1} + {k}}, \label{3}
\ee
where $k$ is an integration constant.

Using \eq{act}, the remaining field equations can be consistently
derived from a truncated action
\be S=-{\tilde N\over \tilde L}\int d\tau \left[(\sinh r)^{m-2}
\Delta-\,\tilde{L}\,(\sinh r)^{m-1} \right]\ ,\label{adsaction}\ee
which gives
\bea
&&\partial_\t \left[\tilde{L}\, ({\sinh r})^{m-2}\, \dot{r}\over
\D \right] = {\cosh r}\,({\sinh r})^{m-3} \left[ (m-1){\sinh r} -
\frac{(m-2)\D}{\tilde{L}} - \frac{\tilde{L}\, (\sinh
r)^2}{\D}\right]
\label{4r}\\
&&\partial_\t \left[ ({\sinh r})^{m-2} \ G_{ab}\
\dot{\f}^b\over \D \, \right] =
\fr{(\sinh r)^{m-2}}{2\D}\,\del_a G_{bc}\,\dot{\f}^b\,\dot{\f}^c\, .
\label{4f}
\eea
In \eq{adsaction}, we introduced $\tilde{N}$ so that
$T_pA_p=\tilde{N}/\tilde{L}^{m-1}$. In particular, for
$(m,n)=\{(5,5),(4,7),(7,4)\}$ we have $\tilde{N}=N$,
$\tilde{N}=\sqrt{N/2}$ and $\tilde{N}=2N^2$, respectively.
Using \eq{3} in \eq{h1}, one finds
\be
E=\tilde{N} k. \label{4en}
\ee
As expected, $k$ is related to $AdS$ energy since it arises from
fixing the time-reparametrization by the static gauge choice $t=\tau$.

To integrate the equations, we introduce flat $(n+1)$-dimensional
coordinates $x^A$ with the constraint $x^A x^A=1$. Then the truncated action
\eq{adsaction} can be rewritten using a Lagrange multiplier
\be
S=-{\tilde N\over \tilde L}\int d\tau \left[(\sinh r)^{m-2}
\Delta-\,\tilde{L}\,(\sinh r)^{m-1}+ L^2 \L (x^Ax^A-1)\right]\ ,\label{adsact2}
\ee
where now $\D^2=\tilde{L}^2\,({\rm cosh}^2{r} - \dot{r}^2)-
L^2\dot{x}^A\dot{x}^A$. The field equations for $x^A$ are
\be
\del_\tau\left[\fr{(\sinh r)^{m-2}\dot{x}^A}{\D}\right]=
-2\L \,x^A,\hs{10} x^A\, x^A=1\, .\label{4xf1}
\ee
By contracting \eq{4xf1}  with $x^A$, it is easy to see that $(\sinh
r)^{m-2}\L/\D=\L_0$,  where $\L_0>0$ is a constant.
The sphere part can be decoupled from the {\it breathing mode} $r$ by
introducing a new time coordinate
\be
d\tilde{\tau}=\fr{\D}{(\sinh r)^{m-2}}\,d\tau .
\ee
In terms of $\tilde{\tau}$, \eq{4f} is equivalent to a
point particle moving on $S^n$. Then, eq. \eq{4xf1} becomes
\be\label{Lagmult}
\fr{d^2x^A}{d\tilde{\tau}^2}=-2\,\L_0\,x^A,
\ee
which can be solved as
\be \label{4sol}
x^A=x_1^A \cos (\sqrt{2\L_0} \tilde{\tau}) + x_2^A \sin(\sqrt{2\L_0}\tilde{\tau}),
\ee
where $x_1^A$ and $x_2^A$ are constants. Imposing $x^A x^A=1$ yields
$x_1^Ax_1^A=x_2^Ax_2^A=1$ and $x_1^Ax_2^A=0$. With these constraints
the total number of integration constants $(x_1^A,x_2^A,\L_0)$ is
$2n$. To determine the moduli space of the solutions, note that
$x_1^A$ defines an $n$-sphere. Being perpendicular to $x_1^A$ and
having unit length, $x_2^A$ defines an $(n-1)$-sphere for each
$x_1^A$. So, the moduli space of the solutions is $R^+$ (corresponding
to $\L_0$) times an $S^{n-1}$ bundle over $S^n$.

The conserved Noether charges for the $SO(n+1)$ invariance of
the action can easily be determined from \eq{adsact2} to be
\be \label{ntvharge}
J^{AB}=\fr{2\L_0\tilde{N}L^2}{\tilde{L}}\left[x_1^Ax_2^B-x_1^Bx_2^A \right].
\ee
Therefore, together with all $[(n+1)/2]$ Cartan generators,
the non-commuting components of $J_{AB}$ can also be activated.

Now let us study the integrability of equations using an explicit
metric on $S^n$. One preferable choice is
\be
d\O_n^2=d\f_1^2+ \cos^2\f_1\,d\f_2^2 +
\sin^2\f_1 \left[d\f_3^2+ \cos^2\f_3^2d\f_4^2+
\sin^2\f_3\left(...+\sin^2\f_{n-1}d\f_{n}^2\right)\right], \label{4sp1}
\ee
so that all Cartan generators of $SO(n+1)$ are manifestly realized as
translations along the cyclic coordinates $\f_2,
\f_4$,..,$\f_n$. Here, the non-cyclic coordinates are defined in the
interval ${[}0,\pi/2{]}$. One may also consider
\be
d\O_n^2=d\f_1^2+\sin^2\f_1\left[d\f_2^2+
\sin^2\f_2(...+\sin^2\f_{n-1}d\f_n^2)\right],
\label{4sp2}
\ee
where $S^n$ is parametrized as nested lower dimensional spheres. Here,
only $\f_n$ is cyclic and others are defined in ${[}0,\pi{]}$. It is also
possible to take combinations of \eq{4sp1} and \eq{4sp2}. Now, \eq{4f}
can be integrated one by one in the order $(\f_n,\f_{n-1},..,\f_1)$
which yields
\be
\dot{\f}_a^2= \frac{\D^2}{L^2(\sinh r)^{2m-4}G_{aa}^2}
\times \left\{ \begin{array}{lll}
q_a^2 \hs{37} \textrm{if $\f_a$ is cyclic} \\ q_a^2 -
\frac{q_{a+1}^2}{{\rm cos}^2\f_a}- \frac{q_{a+2}^2}{{\rm
sin}^2\f_a}\hs{9.5} \textrm{if $\f_a \in{[}0,\pi/2{]}$,} \\
q_a^2-\fr{q_{a+1}^2}{\sin^2\f_a}\hs{23}\textrm{if $\f_a
  \in{[}0,\pi{]}$,}
\end{array}\label{4sphsol} \right.\\
\ee
(no summation is implied in $G_{aa}$) where $q_a$'s  are dimensionless
integration constants with $q_{n+1}=0$.  The motion along the cyclic
coordinates is monotonic.  A non-cyclic coordinate has two turning
points where the time derivatives vanish which are fixed by the
constants $q_a$. The positivity of the velocity-squares imply
\be
q_a\geq q_{a+1}+q_{a+2},\,\,\,\,(\f_a\in{[}0,\pi/2{]})
\hs{5}\textrm{or}\hs{5}q_a\geq q_{a+1}\,\,\,\,(\f_a\in{[}0,\pi{]}).
\label{4b1}
\ee
Above we assumed that all $\f_a$ depend non-trivially on
$\tau$. Otherwise, one has to analyze the field equations to find out
the implications (see below for details).
The metrics \eq{4sp1} or \eq{4sp2} have coordinate singularities and
they are not globally well-defined on $S^n$. Thus \eq{4sphsol} does not cover
the whole solution space and $q_a$'s are not globally well-defined
moduli coordinates unlike the constants $x_1^A$ and $x_2^A$.

In order to exhibit the integrable structure more clearly, one can
express the field equations using the canonical momenta derived from the
action \eq{adsaction}, which gives
\be
P_a^2= \left(\frac{\tilde{N}L}{\tilde{L}}\right)^2 \times
\left\{\begin{array}{lll}
q_a^2 \hs{37} \textrm{if $\f_a$ is cyclic,} \\q_a^2 -
\frac{q_{a+1}^2}{{\rm cos}^2\f_a}- \frac{q_{a+2}^2}{{\rm
sin}^2\f_a}\hs{9.5} \textrm{if $\f_a \in{[}0,\pi/2{]}$,} \\
q_a^2-\fr{q_{a+1}^2}{\sin^2\f_a}\hs{23}\textrm{if $\f_a
  \in{[}0,\pi{]}$.}
\end{array}\right.\\ \label{4sphmom}
\ee
These equations define a canonical transformation
$(\f^a,P_a)\rightarrow (\f^a,q_a)$, such that in the new
variables the equations of motion are $\dot q_{a}=0$ and $\dot{\f}^a$ is
given by \eq{4sphsol}. Clearly, $q_a$'s corresponding to cyclic
coordinates are related to Cartan symmetry generators of
$SO(n+1)$. Other $q_a$'s are  ``hidden'' charges from
the point of view of the sigma model written with the metric \eq{4sp1}
or \eq{4sp2} (for example by evaluating $\D$, one can see that
$q_1^2=2L^2\L_0$).

The potential that appears in \eq{4sphmom} is of
P\"oschl-Teller Type I (for a review see, e.g.,  \cite{ant}). These
belong to a large class, known as shape invariant
potentials \cite{cooper}, which arises naturally in supersymmetric quantum
mechanics and can be solved exactly, i.e. the energy
eigenvalues, eigenfunctions as well as the scattering matrix can
be given explicitly. We shall come back to this later in subsection
\ref{quantizationsph}.

Let us now return to the radial equation which can be fixed using
\eq{delta}, \eq{3} and \eq{4sphsol} as
\be \label{radial}
\fr{1}{2}\dot{r}^2+V(r)=0,
\ee
where
\be\label{p1} V(r)= \fr{\cosh^2 r}{2[(\sinh r)^{m-1}
+k]^2}\left[(q_1^2-k^2)\cosh^2 r\,+\, (k-\sinh^{m-3}r)^2\sinh^2 r
\right].\ee
Eq. (\ref{radial}) is equivalent  to one dimensional motion of a
particle in the effective potential $V(r)$  with zero total energy. In
terms of canonical momenta one has
\be
\fr{1}{2}P_r^2\,+\,\fr{\tilde{N}^2}{2\cosh^2 r}\left[(q_1^2-k^2)\cosh^2
  r\,+\, (k-\sinh^{m-3}r)^2\sinh^2 r \right] =0.\label{prads}
\ee
Unlike the angular part, the potential in \eq{prads} is rather
complicated.

\begin{figure}
\centerline{\includegraphics[width=7.0cm]{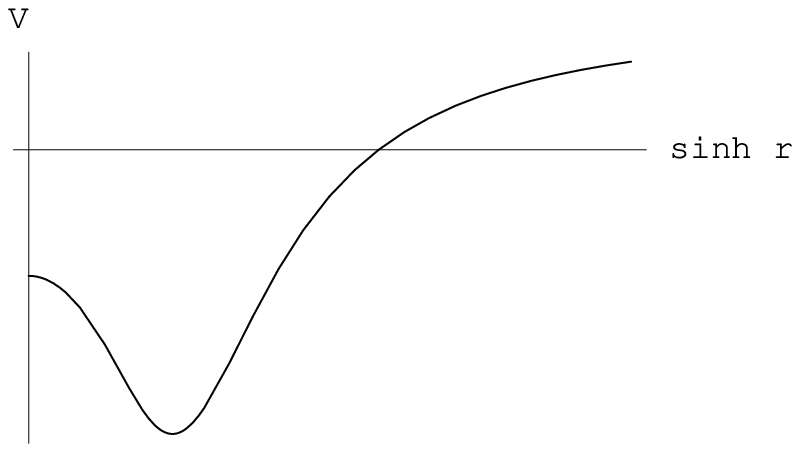}\,\,\,\,
\includegraphics[width=7.0cm]{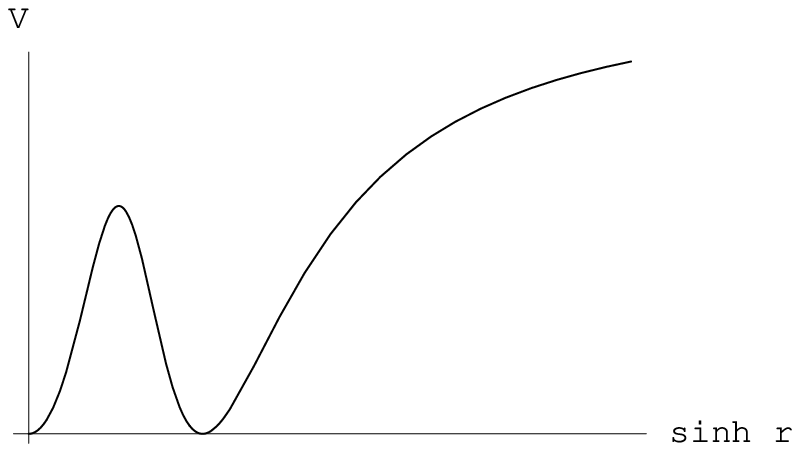}}
\caption{\label{fp1} The potential $V(r)$ in (\ref{p1}) for $k>q_1$ (on
  the left) and for $k=q_1$ (on the right).}
\end{figure}

The motion is allowed in the region $V\leq 0$ which
implies
\be
k\geq q_1.\label{4b2}
\ee
For $k>q_1$, $V=0$ has a single root at
some $r=r_0$ (fixed by $k$, $q_1$ and $m$) and $V>0$ for $r>r_0$ and
$V<0$ for $r<r_0$ (see figure \ref{fp1}). When the brane is given a small
kick at $r<r_0$ (a kick is necessary since the total energy in
this motion is zero), it either climbs the hill and reaches to
$r=r_0$ and rolls back to hit $r=0$ or directly moves through
$r=0$. In either case the brane totally collapses in finite amount
of time and re-expands again. When $k=q_1$, the brane should be
placed at a constant radial distance either at $r=0$ or at $(\sinh
r)^{m-3}=k$ where $V=0$ (see figure \ref{fp1}).

As mentioned above, when a coordinate is set to a constant in the
solution one should check the field equations for its
consequences. These can be summarized as

\

(1) $r=r_0$: From (\ref{radial}) one finds $r_0=0$ or $(\sinh
   r_0)^{m-3}=k$. There is also a third root in (\ref{4r}) with
   $k=(m-2)(\sinh r_0)^{m-3}+(m-3)(\sinh r_0)^{m-1}$ which is not seen
   in figure \ref{fp1}. One should also set $q_1=k$.

\

(2) $\f_a=\f_{0}$ (non-cyclic): This is only possible when
   $\f_a=\{0,\pi/2,\pi\}$; or when $\dot{\f}_b=0$, $b>a$
   for $\f_a\in{[}0,\pi{]}$;  or when all non-cyclic $\dot{\f}_b=0$,
   $b>a$ and the cyclic motion in $b>a$ is confined in a single plane
   defined with $\f_a=\f_0$ for $\f_a\in{[}0,\pi/2{]}$. One can still
   use \eq{4sphsol} by choosing the integration constants
   appropriately (for example if $\f_a\in{[}0,\pi{]}$, then
   $q_{a}\sin^2\f_a=q_{a+1}$.)

\

(3) $\f_a=\f_{0}$ (cyclic): One should set the corresponding
    integration constant to zero.

\

With these taken into account, \eq{4sphsol} and \eq{radial} constitute
our most general configuration. Now, we would like to indicate some special
solutions. One can for example
set $r=r_0$, $\th=0$, $\f_a=\pi/2$ (i.e. all $q_a=0$).
Here, one has to take $r_0\to\infty$ and this is the
``$p$-brane at the end of the universe'' \cite{duff}. Now
if we let $\f_n=\f_n(\tau)$ then we get the 1/2 supersymmetric dual
giant studied in \cite{myers, hashimoto}. in this case, we have
$q_1^2=k^2=q_n^2
=(\sinh r_0)^{2m-6}$  which implies \, $\dot{\f}_n=\tilde{L}/L$.  (This
solution was previously obtained for $AdS_4\times S^7$ in \cite{duff2}
where  $S^7$ is parametrized as a U(1) bundle over $CP^3$.)
A further modification of this configuration is to let $r=r(\tau)$
which was considered in \cite{ouyang}.

Finally, we would like to check whether any of the above solutions
preserve some
supersymmetry. For that purpose we derive a BPS bound for energy whose
saturation is a necessary condition. Actually, using the inequalities
between the integration constants \eq{4b1}, \eq{4b2}, and the conserved
charges \eq{4en}, \eq{4sphmom}, one already finds
\be\label{4bps}
H\geq \fr{\tilde{L}}{L}\sum_a P_a,
\ee
where the sum is over cyclic coordinates. Another way of deriving the
same result is to analyze the Hamiltonian obtained from \eq{adsaction} as
\bea
H=P_{m'}\dot{X}^{m'}-{\cal L}&=&\fr{\tilde{N}\tilde{L}(\sinh r)^{m-2}\cosh^2
  r}{\D}-\tilde{N}(\sinh r)^{m-1},\nn\\
&=& \cosh r\left[\tilde{L}^2 P^2 + \tilde{N}^2(\sinh
  r)^{2m-4}\right]^{1/2}-\tilde{N}(\sinh r)^{m-1},\label{h1}
\eea
where $P^2\equiv P_r^2/\tilde{L}^2\,+\,G^{ab}P_aP_b/L^2$.
Eq. \eq{h1} can be rewritten as
\be\label{trick}
H=\left[(\tilde{L}P+\tilde{N}(\sinh r)^{m-1})^2+(\tilde{L}P\sinh r -
  \tilde{N}(\sinh r)^{m-2})^2\right]^{1/2}-\tilde{N}(\sinh r)^{m-1}.
\ee
This gives $H\geq \tilde{L}P$ and thus
\be
H\geq\fr{\tilde{L}}{L}\sum_a \left[G^{aa}P_aP_a\right]^{1/2},
\ee
where the sum is over cyclic directions. Repeatedly using the inequality
\be\label{id}
\left[\fr{A^2}{\cos^2\th}+\fr{B^2}{\sin^2\th}\right]^{1/2}\geq
A+B\ ,
\ee
one finds \eq{4bps}.

To saturate the bound \eq{4bps}, all other momenta have to vanish. In
this case, a detailed investigation of the field equations shows that
{\it all} rotations have the same angular
velocity and the circular motion is actually confined in a single
plane. This leads us back to the single spin solution after a global
$SO(n+1)$ rotation. Therefore, the only supersymmetric spherical brane
expanding inside $AdS$ is the 1/2  BPS single-spin giant graviton.

\subsection{Branes Expanding in Sphere (Magnetic)}

In this case the background supports an $n$-form field strength given by
$H_{\hat{\theta}\hat{\phi}\hat{\chi}_1...\hat{\chi}_{n-2}}=(n-1)/L$
in the orthonormal basis (see footnote \ref{ft}). We rewrite (\ref{metric}) as
\be ds^2= \tilde{L}^2\, G_{\m\n}dy^{\m}dy^{\n}+ L^2 d\O_n^2,
\hs{7}\m,\n=0,..,m-1,
\label{4ads}
\ee
where $G_{\m\n}$ is the metric on the unit $AdS_m$, $y^{0}=t$ and the
sphere parametrization is given in (\ref{sp1})
\be
d\O_n^2 =
d\th^2 +{\rm cos}^2\th d\f^2+{\rm sin}^2\th\left[d\chi_1 ^2+ {\rm
sin}^2\chi_1 (...+ {\rm sin}^2\chi_{n-3}d\chi_{n-2} ^2)\right]\ .
\label{4sphere}
\ee
We identify the world-volume coordinates with
\be t= \s_0 = \t, \hs{5}
\chi_1=\s_1,\,\,...,\,\, \chi_{n-2}=\s_{n-2},
\ee
and assume a solution of the form:
\be\label{an2}
\f=\f(\t), \hs{5} \th=\th(\t), \hs{5} y^\m=y^{\m}(\tau).
\ee
The induced metric becomes
\be \g_{\a\b}=\partial_{\a}X^M\partial_{\b}X^N g_{MN}=
\left( \begin{array}{cc} - \D^2 & 0 \\
0 & L^2\, {\rm sin}^2\theta \ (g^\chi)_{mn}\end{array} \right), \ee
where $(g^\chi)_{mn}$ is the metric on the unit $(n-2)$-sphere and
\be \D^2= -\tilde{L}^2\,(G_{\m\n}\dot{y}^{\m}\dot{y}^{\n})
-L^2\,(\dot{\th}^2+\dot{\f}^2{\rm cos}^2\th).
\label{delta2}
\ee
Eq. (\ref{field}) is satisfied trivially for $M=\chi_i$, and $M=t$ component
fixes the on-shell value of $\D$ as
\be
\D = k\, \tilde{L}\, ({\rm sin}\theta)^{n-2} \, {\cosh}^2 r, \label{sd}\\
\ee
where $k$ is an integration constant.

Using \eq{act}, the remaining equations can be obtained from the
following one dimensional truncated action
\be S=-{N\over L}\int d\tau \left[(\sin \th)^{n-2}
\Delta-L\,\dot{\f}\,(\sin\th)^{n-1}\right]\ ,\label{4sact} \ee
which gives
\bea
&&\partial_\t \left[L\,({\sin \th})^{n-2} \cos^2 \th \,\dot{\f}\over \D
\right] =-\del_\tau (\sin\th)^{n-1},\label{s11}\\
&&\partial_\t \left[L\,({\sin \th})^{n-2}\,\dot{\th}\over \D \right] = \cos\th
(\sin\th)^{n-3}\left[(n-1)\sin\th\,\dot{\phi}-\fr{(n-2)\D}{L}-
\fr{L\,\sin^2\th}{\D}\dot{\phi}^2\right],\label{s2}\\
&&\partial_\t \left[ (\sin \th)^{n-2} \, G_{\m\n}\
\dot{y}^\n \over \D \, \right] =  \fr{(\sin\th)^{n-2}}{2\D}\,\del_\m\,
G_{\n\rho}\,\dot{y}^{\n}\dot{y}^{\rho}. \label{s3}
\eea
Using the on-shell value of $\D$ \eq{sd} in \eq{ham3} one gets
\be\label{staticads}
E=\fr{N\tilde{L}}{kL}.
\ee
As before the integration constant $k$ is related to $AdS$ energy.

Similar to the previous subsection, one can introduce a new world-volume
time coordinate as follows:
\be
d\tilde{\tau}=\fr{\D}{(\sin\th)^{n-2}}\,d\tau.
\ee
In terms of $\tilde{\tau}$, the {\it breathing mode} $\th$ and $\f$
coordinate are decoupled from the rest and \eq{s3} corresponds to a
point particle moving in $AdS_m$. On the other hand, \eq{s11} and
\eq{s2} give motion on $S^2$ with an additional velocity dependent
potential.

Let us first analyze the $AdS$ part. If $Y_P$ are $(m+1)$-dimensional
embedding coordinates, then the  $AdS$ space is defined by
$\eta^{PQ}Y_{P}Y_{Q}+1=0$, where $\eta_{PQ}=(-1,-1,+1,..,+1)$. We take
all $Y_P$ to be dynamical (namely, we are not imposing the static
gauge). Using a Lagrange multiplier, the action \eq{4sact} can be rewritten as
\be S=-{N\over L}\int d\tau \left[(\sin \th)^{n-2}
\Delta-L\,\dot{\f}\,(\sin\th)^{n-1}+\tilde{L}^2\L(\eta^{PQP}Y_PY_Q+1)\right].
\ee
where now $\D^2=
-\tilde{L}^2\,\eta^{PQ}\dot{Y}_P\dot{Y}_Q-L^2\,(\dot{\th}^2+\dot{\f}^2{\rm
  cos}^2\th)$ and dot again denotes differentiation with respect
to $\tau$. Varying for $Y_P$ and $\L$ one obtains
\be\label{adsgl}
\del_\tau\left[\frac{(\sin\th)^{n-2}\,\dot{Y}^{P}}{\D}\right]=-2\L\,Y^{P},
\hs{5}
\eta^{PQ}Y_PY_Q=-1.
\ee
Contracting \eq{adsgl} with $Y_P$, one finds
$(\sin\th)^{n-2}\L/\D=\L_0$, where $\L_0$ is an arbitrary real
number. Then \eq{adsgl}  becomes
\be\label{yeqn}
\fr{d^2Y^P}{d\tilde{\tau}^2}=-2\,\L_0\,Y^P,
\ee
which can be integrated in terms of elementary functions. The solution
space consists of three disjoint parts parametrized by $\L_0<0$,
$\L_0=0$ and $\L_0>0$. The global $SO(2,m)$ charges can be calculated
as
\bea
S_{PQ}&=&\fr{\tilde{L}^2N(\sin\th)^{n-2}}{L\D}\left(Y_P\dot{Y}_Q-Y_{Q}
\dot{Y}_P\right)\nn\\ &=&
\fr{\tilde{L}^2N}{L}\left(Y_P\del_{\tilde{\tau}}Y_Q-Y_{Q}
\del_{\tilde{\tau}}Y_P\right),
\eea
which is clearly conserved by \eq{yeqn}. Note that the $AdS$ energy
obtained in this way will in general differ from \eq{staticads}, since
the later was calculated in the static gauge.

It is interesting to analyze the integrability of the  equations using
local coordinates on $AdS$
\be\label{ads33}
G_{\m\n}dy^\m dy^\n=-\cosh^2r\, dt^2 + dr^2 + \sinh^2 r\, G_{ij}\, d\a^i
d\a^j,
\ee
where $G_{ij}$ is the metric on unit $S^{m-2}$. One can solve the
equations \eq{s3} starting from the spherical part which is
essentially the same problem worked in the previous subsection. Using
the parametrizations given in \eq{4sp2} or \eq{4sp1}, the result is
\be
\dot{\a}_i^2 = \frac{{k}^2\, {\rm cosh}^4{r}}
{G_{ii}^2\sinh^4 r} \times \left\{\begin{array}{lll}
q_i^2 \hs{37} \textrm{if $\a_i$ is cyclic} \\ q_i^2 -
\frac{q_{i+1}^2}{{\rm cos}^2\a_i}- \frac{q_{i+2}^2}{{\rm
sin}^2\a_i}\hs{9.5} \textrm{if $\a_i \in{[}0,\pi/2{]}$} \\
q_i^2-\fr{q_{i+1}^2}{\sin^2\a_i}\hs{23}\textrm{if
  $\a_i\in{[}0,\pi{]}$},
\end{array}\right.\\
\label{alpha}
\ee
(no summation in $G_{ii}$) and
\be
\dot{r}^2= {\rm cosh}^4{r}\, \left[- q^2- \frac{q_1^2\,
k^2}{{\rm sinh}^2{r}} + \frac{1}{{\rm cosh}^2{r}}
\right] \label{radial2}
\ee
where $q$ and $q_i$ ($q_{m-1}=0$)  are dimensionless
constants of motion which should obey
\bea
&&1\geq q_1\,k+q, \nn\\
&&q_i\geq
q_{i+1}+q_{i+2},\,\,\,\,(\a_i\in{[}0,\pi/2{]})\hs{5}\textrm{or}
\hs{5}q_i\geq q_{i+1}\,\,\,\,(\a_i\in{[}0,\pi{]}).
\label{bpsin1}
\eea
The non-cyclic $\a_i$ and  the coordinate $r$ pulsate between two
turning points fixed by $q$ and $q_i$.
The canonical momenta for the angular coordinates $\a_i$ are given by
\be
P_i^2= \left(\frac{N\tilde{L}}{L}\right)^2 \times
\left\{\begin{array}{lll}
q_i^2 \hs{37} \textrm{if $\a_i$ is cyclic,} \\q_i^2 -
\frac{q_{i+1}^2}{{\rm cos}^2\a_i}- \frac{q_{i+2}^2}{{\rm
sin}^2\a_i}\hs{9.5} \textrm{if $\a_i \in{[}0,\pi/2{]}$,} \\
q_i^2-\fr{q_{i+1}^2}{\sin^2\a_i}\hs{23}\textrm{if $\a_i
  \in{[}0,\pi{]}$,}
\end{array}\right.\\ \label{pais}
\ee
On the other hand \eq{radial2} become
\be
P_r^2= \left({N\tilde L\over kL}\right)^2
\left[-q^2-{q_1^2k^2\over\sinh^2r}+{1\over\cosh^2
r}\right]\ .\label{prs}
\ee
The potential for $P_r$ in \eq{prs} is of P\"oschl-Teller Type II
\cite{cooper}.

Returning to the $\th$ and $\f$ coordinates, we see that \eq{s11} readily
determines $\f$
\be
\dot{\phi} = \frac{k\, \tilde{L}\, {\rm cosh}^2{r}}{L\,
{\rm cos}^2\theta}\,
[p-({\rm sin}\theta)^{n-1}],  \label{delta3}
\ee
where $p$ is an integration constant. On the other hand, $\th$ can be
fixed from \eq{delta2} and \eq{sd} which yields
\be
\fr{1}{2}\,\dot{\theta}^2 +V(\th)=0, \label{theta4}
\ee
where
\be \label{pot2} V(\th)= \frac{\tilde{L}^2\, {\rm
    cosh}^4{r}}{2L^2}\,\left[k^2p^2-q^2+\fr{k^2\sin^2
    \th}{\cos^2\th}(p-\sin^{n-3}\th)^2\right].
\ee
This is a one-dimensional motion in the potential $V(\th)$ with zero
total energy. The requirement $V(\th)\leq 0$ implies
\be\label{bpsin2}
q\geq kp.
\ee
When expressed in terms of momenta, \eq{delta3} and \eq{theta4} become
\bea
&&P_\f=Np,\label{pais2}\\
&&P_\th^2= {N^2\over
k^2}\left[k^2p^2-q^2+\fr{k^2\sin^2
    \th}{\cos^2\th}(p-\sin^{n-3}\th)^2\right].
\label{pths}
\eea
The separation in canonical variables is manifest in \eq{pais},
\eq{prs}, \eq{pais2} and \eq{pths}. Unlike the
exactly solvable potentials we encountered above, the potential on the
right hand side of \eq{pths} is more complicated.

\begin{figure}
\centerline{\includegraphics[width=7.0cm]{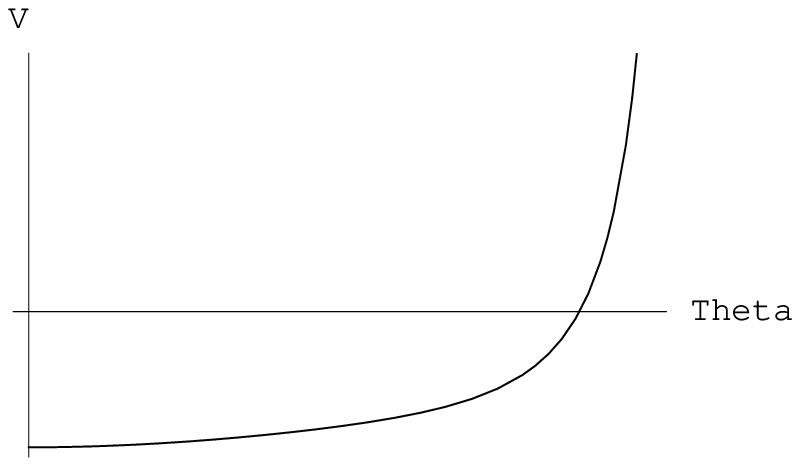}\,\,\,\,
\includegraphics[width=7.0cm]{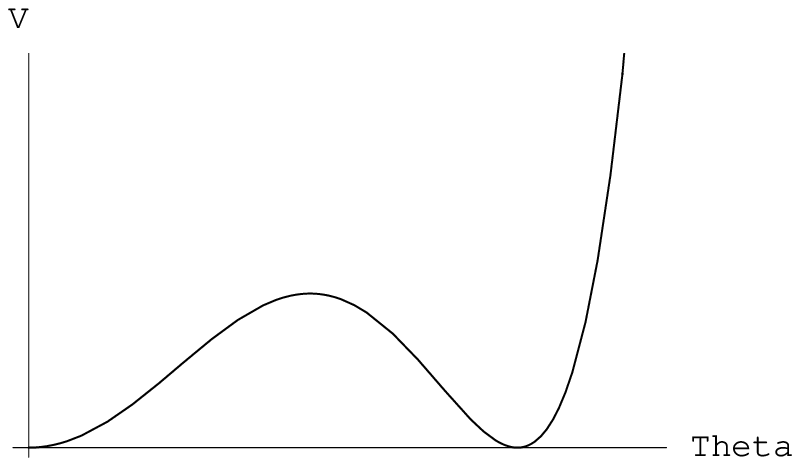}}
\caption{\label{fp2} The potential $V(\th)$ in (\ref{pot2}) for
  $q<kp$ (on the left) and for  $q=kp$ (on the right).}
\end{figure}

Let us now discuss the motion for the breathing coordinate $\th$ in
more detail.
When $q<kp$, $V(\th)$ is nowhere negative
or zero.  For $q=kp$, $V(\th)=0$ only at $\th=0$ and at $({\rm
sin}\theta)^{n-3}=p$ and $V(\th)>0$ otherwise (see figure \ref{fp2}). This
enforces the brane to locate at either root and note that the
second zero exists only for $p\leq1$. When $q>kp$ there are two
possibilities. Firstly, if $p=1$ $V(\th)$ is always negative and
in this case $\th$ reaches $0$ or $\pi/2$ in a finite amount of time.
Secondly, if $p\not=1$ there is a root $\th_0$ so that
$V(\th)\leq0$ for $\th\leq\th_0$ and $V(\th)>0$ for $\th>\th_0$
(such a potential is drawn in figure \ref{fp2}). In this case, the brane
either contracts  directly to $\th=0$ or it expands till
$\th=\th_0$ and then collapses to $\th=0$ in a finite amount of
time. Note that $V(\th)$ is scaled by $\tilde{L}^2\cosh^4r/L^2$
which does not alter the zeroes but affect the shape of the
potential when $r$ is time dependent.

Equations \eq{alpha} and \eq{radial2} hold under the assumption that
the coordinates have non-trivial time dependences.  Otherwise the
original field equations may impose further restrictions. These
can be summarized as:

\

(1) $\f=\f_0$: From (\ref{s11}) this happens only when $\th=\th_0$
and from (\ref{s2}) one has $\th_0=0, \pi/2$. The constant $p$ should
be fixed as $p=1$.

\

(2) $\th=\th_0$: This requires $q=kp$ and from (\ref{s3}) one
finds $({\rm sin}\theta_0)^{n-3}=p$, \,$\th=0$,\, or
$\th=\pi/2$. There is also a fourth root with
$p=(n-2)(\sin\th_0)^{n-3}-(n-3)(\sin\th_0)^{n-1}$.

\

(3) $r=r_0$: From (\ref{s2}) and (\ref{radial2}) we see that
  $q^2=1/\cosh^4 r_0$\,  and\,  $q_1 k=\sinh^2 r_0/\cosh^2 r_0$.

\

(4) $\a_i=\a_0$: This is exactly the same with the special cases (2)
    and (3) discussed in the previous subsection.

\

Upto these four cases, \eq{alpha}, \eq{radial2}, \eq{delta3}
and \eq{theta4} give the most general configuration.
The 1/2 supersymmetric two-spin solution (\ref{sol}) is  obtained when we have
$\a_1=...=\a_{m-3}=\pi/2$\, and\, $\theta=\th_0$, \,\,$r=r_0$.
Using the above conditions one finds $q_1=...=q_{m-2}$,
$p=(\sin\th)^{n-3}$, $q=kp=1/\cosh^2r$ and $q_1k=\sinh^2/\cosh^2r$
which imply from (\ref{delta3}) and (\ref{alpha}) that
$\dot{\phi}=\tilde{L}/L$\, and \, $\dot{\a}_{m-2}=1$ (recall that
here $t$ is dimensionless). If we further set $\a_{m-2}=\pi/2$,
then $q_1=..=q_{m-2}=0$ which gives $r=0$. This is the 1/2
supersymmetric giant graviton solution of \cite{susskind}.  For
this case, $\th=\th(\tau)$ is studied in \cite{ouyang}. There is
also the trivial solution with $r=0$, $\th=\pi/2$ and all other
coordinates are set to constants.

Let us  conclude this subsection by  deriving a BPS bound  for the energy
given the  conserved charges. This  can readily be obtained  using the
inequalities   between    the   integration   constants   \eq{bpsin1},
\eq{bpsin2},   and    the   expressions   for    conserved   quantities
\eq{staticads},  \eq{pais},  \eq{pais2}  which  gives
\be\label{bps4}
H\geq \fr{\tilde{L}}{L}  P_\f+\sum_i P_{i},\ee
where the sum  is over cyclic coordinates. To obtain the  same result
in a different way, the Hamiltonian can be found from \eq{4sact}  as
\bea
H=P_{m'}\dot{X}^{m'}-{\cal L}
&=&\fr{\tilde{L}^2N(\sin\th)^{n-2}\cosh^2r}{L\D},\nn\\
&=&\tilde{L}\cosh r
\left[P^2+\fr{N^2}{L^2}(\sin\th)^{2n-4}\right]^{1/2}, \label{ham3}
\eea
where
$P^2\equiv P_\f^2/(L^2\cos^2\th)+G^{\m\n}P_\m P_\n/L^2$. The
last    term    in    (\ref{ham3})  can be    combined    with
$P_\f^2/(L^2\cos^2\th)$ to  give $P_\f^2/L^2$ plus an  exact square as
in (\ref{ham1}).  Then, by  a calculation similar  to the one  done in
(\ref{b1}),  and  using  the  inequality  (\ref{id})  repeatedly,  the
Hamiltonian (\ref{ham3}) can be shown to subject to \eq{bps4}.

This BPS bound can only be realized when all other momenta except
$P_\f$ and cyclic $P_i$ vanish which enforce the corresponding
coordinates to be constants. However, as in the previous subsection,
one can see  that the circular motion in $S^{m-2}$ is along the
equator,  which is equivalent to the 1/2 supersymmetric two-spin
solution. This reduces to the single-spin case when the radius
of the embedded $S^{m-2}$ goes to zero.

\subsection{Quantization of Spherical Giants}

\label{quantizationsph} We have examined spherically symmetric $p$-branes
in $AdS_m\times S^n$, which are defined by \eq{an1} in the electric case
($p=m-2$) and \eq{an2} in the magnetic case ($p=n-2$). At the quantum
level, the spherical truncation is performed by first fixing a physical
gauge and expanding in normal coordinates $\xi^{m'}$ defined by
$X^M(\sigma^\mu)=\left(\s^\mu,X^{m'}(\tau)+\xi^{m'}(\s^\mu)\right)$,
where $X^{m'}(\tau)$ describe the 0+1 dimensional sigma-model. The
classical consistency of the truncation implies that the action has no
linear terms in $\x^{m'}$. For large brane tension, the normal
coordinates $\xi^{m'}$ become free, and thus yield a one-loop determinant
contribution to the 0+1 dimensional sigma model for $X^{m'}$. We shall
omit this contribution, based on the fact that the free spectrum for
$\xi^{m'}$ is evenly spaced both for the single-spin and the two-spin solutions
and only depends on the $AdS$ and sphere radii. In this approximation, the
spherical giant is described by a wave-function $\psi(X^{m'})$, where
$X^{m'}$ are the transverse coordinates and the canonical momenta are
realized by $P_{m'}=-i\partial/\partial X^{m'}$.

In the classical theory, the spherical brane consists of a warped product
of a breathing-mode and a relativistic, massive point-particle. The
latter lives on $R\times S^n$ in the electric case, where $R$ is time,
and on $AdS_m\times S^1$ in the magnetic case, where $S^1$ is the cyclic
direction in $S^n$ transverse both to the brane and the breathing
direction. We have found that the resulting 0+1 dimensional sigma model
is an integrable system\footnote{See \cite{min} in the case of the string
in $AdS_5\times S^5$. Also, non-relativistic point particles on spheres
with potential appear in a similar context, namely as truncation of the
string sigma-model, corresponding to picking a particular solution to
conformal gauge, leading to the soluble N-R model, see e.g.
\cite{tseytlin4}, \cite{NR1} and \cite{NR2}.}, and that the breathing
mode is governed by a potential that depends on the total angular
momentum of the point-particle.

The quantization can be performed in terms of global embedding
coordinates. This is discussed in Appendix \ref{ap1}. In this subsection,
we shall instead quantize using local spherical or $AdS$ coordinates,
leading to P\"oschl-Teller potentials. In doing so, we parametrize the
spheres using {\it maximal number of cyclic coordinates}, i.e. \eq{4sp1},
and  impose \eq{4sphmom} in the electric case, and \eq{pais}, \eq{prs},
and \eq{pais2} in the magnetic case. We then proceed with the breathing
modes, using the results of \cite{ouyang}.

The classical solutions are parameterized by constants of motion, namely
$(k,q_a)$ defined in \eq{4sphmom} and \eq{prads} in the electric case,
and $(k,q,q_i)$ defined in \eq{pais}, \eq{prs} and \eq{pths} in the
magnetic case, and one should ask whether they are actually limits of
states in the quantum theory. To begin with, it follows from their
definition that their mutual Poisson brackets vanish. The constant $k$
determines the $AdS$ energy $E$. The $AdS$ energy, which is identified as
the $p$-brane Hamiltonian, is quantized since the spatial world-volume is
compact, though in the absence of additional symmetries there is no
mechanism preventing non-integer energies. To be more precise, $k$ sets
the energy levels of the potential for the breathing modes given in
\eq{prads} and \eq{pths}, which become discrete in the quantum theory.

The point-particle motion may therefore be thought of as an internal
sector, analogous to the orbital angular momentum in a central force
problem. In this sector the Cartan subalgebra generators of
$SO(m-1)\times SO(n+1)$ that are not set to zero in the spherical
reduction are determined by some of the $q_a$ and $(q,q_i)$, namely the
momenta of the cyclic transverse spatial coordinates. These symmetries
are realized in the world-volume quantum theory, in the limit where this
theory becomes reliable. Hence, the non-vanishing Cartan subalgebra spins
are integers, which we shall denote by $S_i=P_i=n_i\subset SO(m-1)$ and
$J_i=P_i=n_i\subset SO(n+1)$

The remaining $(q_a)$ and $(q,q_a)$ are activated by the oscillatory
point-particle motion in $AdS_m\times S^n$, i.e. they set energy-like
levels for the P\"oschl-Teller potentials, which become quantized with
discrete spectra (there is also a continuum in the Type II potential).
This quantization can also be understood as quantization of the Casimirs
of the chains $SO(m-1)\supset SO(m-3)\supset\cdots \supset
SO(m-1-2[(m-4)/2])$ and $SO(n+1)\supset SO(n-1)\supset\cdots\supset
SO(n+1-2[(m-2)/2])$, arising in the parametrization of the sphere line
elements using maximal number of cyclic coordinates.\footnote{We thank K.
Murakami for this observation.}

The P\"oschl-Teller potentials belong to a large class of exactly
solvable quantum mechanical models, defined by superpotentials with a
special property known as shape invariance \cite{cooper}. There is a
group-theoretic approach to solving the P\"oschl-Teller potentials, based
on coset representations of $S^2$ and $AdS_2$ (see, e.g., \cite{ant}). In
Appendix \ref{ap0}, we summarize its energy spectrum.

From the parametrization of the Type I potentials given
in \eq{4sphmom} and \eq{pais} and using \eq{type1}  one finds
($\phi_i$ oscillatory):
\be T\,q_i=1
+\sqrt{\fr14+(T\,q_{i+1})^2}+\sqrt{\fr14+(T\,q_{i+2})^2}+ 2n_i\
,\label{Tai}\ee
where $T=L{\tilde N}/{\tilde L}$ (electric) and $T={\tilde L} N/L$
(magnetic). (To compare with WKB approach see Appendix \ref{ap2}).
Note that $T\, q_{i+1}$ is a spin, and also $T\,q_{i+2}$ is a spin in
case $i$ is the last oscillatory coordinate.

In the electric case we define $Q=T\,q_1$, and the above formula
gives
\be Q=J_1+\cdots J_v+2(n_1+\cdots+n_{n-v})+\fr{(n-1)}{2}+{\cal O}(1/T),
\label{bspt1}
\ee
where $J_i$ denote the spins in the Cartan subalgebra, and
$v=[(n+1)/2]$.

In the magnetic case the remaining point-particle motion in $r$ is
governed by the Type II potential given in \eq{prs}. From \eq{type2},
the discrete spectrum is given by
\be - Q = 1+Tq_1-E+2n_r+{\cal O}(1/T)\ ,\qquad 0\leq n_r\leq
(E-Tq_1)/2,\label{bspt2}\ee
where $Q=Tq/k$, $E=N\tilde L/(kL)$ (and we have chosen the sign
leading to a positive contribution to $E$), and
\be Tq_1=S_1+\cdots +S_v+2(n_1+\cdots+n_{m-1-v}) + \fr{(m-3)}{2}+{\cal
  O}(1/T) \ ,\label{bspt3}\ee
where $v=[(m-1)/2]$. As one approaches the continuum of the Type
II potential, $Q$ decreases, which leads to that the breathing
mode becomes strongly coupled. This is an interesting region, as
the $p$-brane starts probing large $r$, but for simplicity we
shall continue under the assumption $n_r\ll (E-Tq_1)$.

The quantization of the breathing modes, i.e. $r$ and $\th$ are
governed by \eq{prads} and \eq{pths} in the electric and magnetic
cases, respectively, was studied separately in \cite{ouyang},
using semi-classical techniques, and we shall review these results
in the present context below

Here the
wave-function is approximated by the exponential of the action
integral, and the boundary conditions are approximated by the
Bohr-Sommerfeld quantization conditions
\bea
I_r &=&\oint\, dr\, P_r = 2\pi (n_r+\fr12)\nn \\
&=&\tilde N\oint {dr\over \cosh r}
\left[(k^2-q_1^2)\cosh^2 r\,-\, (k-\sinh^{m-3}r)^2\sinh^2 r
\right]^{1/2}\, , \\
I_\th&=&\oint\,d\th\, P_\th=2\pi(n_\th+\fr12)\nn\\
&=&\fr{N}{k}\oint d\th
\left[q^2-k^2p^2-\fr{k^2\sin^2
    \th}{\cos^2\th}(p-\sin^{n-3}\th)^2\right]^{1/2}.\eea
This approximation method is good if $1\ll n\ll N$, where the upper
bound is set such that the total $AdS$ energy will not become too
large, while the lower bound is set so that the WKB approximation
can be trusted. Clearly, only the upper bound is necessary from
the spacetime point of view, while the lower bound is simply an
artifact of the particular approximation method used to solve the
Schr\"odinger problem. The upper bound implies that it makes sense
to expand in $n/T$, where $T$ is the tension defined under
\eq{Tai}. Adapting the results of \cite{ouyang} to our cases, the
quantization of $r$ in the electric case with $(E-\tilde{L}Q/L)
\ll \tilde N, \, Q$ gives (up to $n_r^3$ terms)
\be E-\frac{\tilde L}{L}Q= \frac{2\tilde L}{L}(n_r+\fr12)-
\left\{\begin{array}{ll}{6Nn_r^2 \over
Q^2}\ ,&(m,n)=(5,5), \hs{5} N=\tilde N\\
\frac{12Nn_r^2}{Q^3},\, &(m,n)=(4,7), \hs{5} N=2\tilde N ^2
\\\frac{15Nn_r^2}{2Q^{3/2}},\, &(m,n)=(7,4). \hs{5}
N=\sqrt{\frac{\tilde N}{2}}
\end{array}\right.\\
\label{energyquant}\ee
The $\th$ quantization in the magnetic case with
$(Q-\tilde{L}P_\phi/L) \ll N$ gives
\be Q -\,\fr{\tilde{L}}{L}\,P_\phi=2(n_\th+\fr12)-
\left\{\begin{array}{ll}{6Nn_\th^2 \over P_\phi^2}\ ,&(m,n)=(5,5),\\
\frac{15N^{1/2}n_\th^2}{2 P_\phi^{3/2}},\, &(m,n)=(4,7), \\
\frac{6N^2n_\th^2}{P_\phi^3},\, &(m,n)=(7,4),
\end{array}\right. \\ \label{ou2}
\ee Compared to \eq{Tai}, here we have a series expansion in $n$ which
makes one wonder whether the Bohr-Sommerfeld method used in \cite{ouyang}
would reproduce the exact result \eq{Tai} for P\"oschl-Teller type
potential. As we show in Appendix \ref{ap2}, this indeed happens up to
fourth order in $n$ but $1/T$ corrections that are present in \eq{bspt1},
\eq{bspt2} and \eq{bspt3}  are not observed.

\subsection{Holography}
\label{hol} The above results can be summarized by the following
expressions for the $AdS$ energy in the case of electric and magnetic
spherical $p$-branes for $(m,n)=\{(5,5),(4,7),(7,4)\}$:
\bea E_{\rm el}&-&\fr{\tilde{L}}{L}\left[J_{1}+\cdots
  +J_{v}\right]=\fr{\tilde{L}}{L}\left[2
(n_r+n_1+\cdots+n_{n-v})+\fr{(n+1)}{2}\right]\nn\\
&-&\fr{(n-1)(n+1)N}{2^{[(n+3)/4]}Q^{(n-1)/2}}n_r^2+{\cal
O}(n_{r}^3)+{\cal O}(1/\tilde N)\ ,\label{Eel}\\
E_{\rm mag}&-&\left[S_1+\cdots
+S_v\right]-\fr{\tilde{L}}{L}P_\phi=
2(n_\th+n_r+n_1+..+n_{m-1-v})\nn\\&+&\fr{(m+1)}{2}
-\fr{(m-1)(m+1)N^{(m-3)/2}}{2^{[(m-1)/2]}P_\phi^{(m-1)/2}}n_\th^2+{\cal
O}(n_{\th}^3)+{\cal O}(1/N), \label{Emagn}\eea
where $v=[(n+1)/2]$ (electric) and $v=[(m-1)/2]$ (magnetic). The finite
$n$ and $m$ dependent shifts should cancel in the supersymmetric
completions (in order not to violate the unitarity bounds imposed by the
superalgebra). The above expressions for the $AdS$ energy are valid for
bosonic $p$-branes. They may be altered in the case of supersymmetric
$p$-branes due to contributions from the fermions, since in the quantum
mechanical Hamiltonian the fermions cannot be truncated.

Following the proposal of \cite{berkooz1,jevicki}, we assume that a bulk
$D3$-brane wave-function(al) with energy $E$, compact $SO(4)\subset
SO(4,2)$ spins $(S_1,S_2)$, and $SO(6)$ Cartan spins $(J_1,J_2,J_3)$,
corresponds to a component of an operator ${\cal
O}_{[R](\D;S_LS_R;m_1m_2m_3)}$ in the dual SYM theory with scaling
dimension $\D=E$, non-compact $SL(2,C)$ (Lorentz) spin $(S_LS_R)$, and
$SO(6)$ highest weights $(m_1m_2m_3)$. Note that under $SO(6)\rightarrow
U(1)^3$ the operator decomposes into components with Cartan spins
$(J_1,J_2,J_3)$ obeying $|J_1|+|J_2|+|J_3|=m_1$. The index $[R]$ refers
to a representation of $S_N$, i.e. a Young-tableau of size $k$, used to
construct the $SU(N)$ invariant \cite{jevicki}. The two extreme cases are
single-column Young-tableaux, i.e. subdeterminants, which correspond to
magnetic $D3$-branes, and single-row Young-tableaux which correspond to
electric $D3$-branes. In these cases, an operator of size $k$ has the
form
\be {\cal O}_{[\pm]}={\cal O}_{(\l_1)\dots(\l_k) } \d^{i_1\dots
i_k}_{\{j_1\dots j_k\}_{\pm}} W^{(\l_1)}{}_{i_1}^{j_1}\cdots
W^{(\l_k)}{}_{i_k}^{j_k},\label{calo1}\ee
where $\{\cdots\}_{+}$ and $\{\cdots\}_-$ denote symmetrization and
anti-symmetrization, respectively, and $W^{(\l)}$ denote derivatives of
the $SU(N)$ valued SYM superfield $X^A$. The tensor ${\cal O}_{\l_1\dots
\l_k}$ picks out some irrep and contains an $N$-dependent normalization
of two-point function. Two basic properties of these type of operators
are that 1) they form a diagonal set for large $k\sim N$, which is
possible because they contain multi-trace contributions which are not
suppressed for large $N$; and 2) the constituents are automatically
symmetrized (for both $\pm$). Consider a fixed ${\cal O}_{0}$ of size
$k\sim N$, $N-k\gg 1$, and the space of excitations built on ${\cal
O}_{0}$ by inserting impurities while keeping $R$ in single column or row
form. The above properties imply a Fock space structure
\cite{berkooz2,berenstein}, which is isomorphic to the space of
multi-particle states (i.e. composites) of the super Maxwell field theory
arising from normal-coordinate expansion around the semi-classical
$D3$-brane giant solution. In particular, the super Maxwell ground state
carries the same charges as ${\cal O}_{0}$.

Let us consider the wave function of a spherically symmetric electric
$D3$-brane giant with quantum numbers $(J_1,J_2,J_3;n_1,n_2;n_r)$. The
corresponding operator ${\cal O}_{(J_1,J_2,J_3;n_1,n_2;n_r)}$ is a
Lorentz scalar with $SO(6)$ Cartan spins $(J_1,J_2,J_3)$. Eq. \eq{Eel}
suggests that the bare scaling dimension is given by
\be \D_{\rm bare} =J_1+J_2+J_3+2(n_1+n_2+n_r)=Q+2n_r,\ee
which implies that for $n_r$ the operator ${\cal
O}_{(J_1,J_2,J_3;n_1,n_2;0)}$ is protected. Hence, the set of operators
\be {\cal O}_{(J_1,J_2,J_3;n_1,n_2;0)}\ ,\qquad J_1+J_2+J_3=J,\qquad
J+2(n_1+n_2)=Q,\label{setofO}\ee
may be identified with the components of the protected scalar
chiral primary operator ${\cal O}_{(J;00;J00)}$:
\be {\cal O}_{(J;00;J00)}={\cal O}_{A_1\dots A_J}\d^J X^{A_1}\cdots
X^{A_J}\label{chiralO}\ee
where $A$ is the $SO(6)$ vector index, $X^A$ the $SU(N)$ valued
singletons and ${\cal O}_{A_1\dots A_J}$ is a constant traceless $SO(6)$
tensor, and $\d^J$ denotes the symmetric $SU(N)$ invariant of size $J$ in
\eq{calo1}. Under $SO(6)\rightarrow U(1)^3$, the singleton superfields
decompose as $X^A\rightarrow (Z_1,Z_2,Z_3)$, where $Z_i=X^{2i-1}+i
X^{2i}$. If one lets
\be K_{\a_1,\a_2,\a_3}=\a_1|Z_1|^2+\a_2|Z_2|^2+\a_3|Z_3|^2\ ,\ee
then ${\cal O}_{(\D;00;J00)}$ decomposes into the set of
components
\be \d^J Z_1^{J_1} Z_2^{J_2} Z_3^{J_3}
K_{2,-1,-1}^{l_1}K_{-1,2,-1}^{l_2}\ ,\qquad J_1+J_2+J_3=J\ ,\qquad
J+2(l_2+l_2)=Q,\label{setofO2}\ee
where $K_{2,-1,-1}$ and $K_{-1,2,-1}$ is a choice of basis for
traceless, $U(1)^3$-invariant bilinears. Elementary counting shows
that the two sets of operators \eq{setofO} and \eq{setofO2} are
isomorphic. The identification becomes manifest in global
coordinates, where the giant wave-function is given by
\eq{elecwave} for $p=0$.

From \eq{Eel}, we see that starting from the protected operators making
up the components of ${\cal O}_{(J;00;J00)}$, and switching on finite
breathing number, $n_r=1,2...$, adds a bare dimension $2n_r$ and a
negative anomalous dimension, $-6Nn_r^2/J$. Finite breathing number
implies that the giant wave-function depends on the $AdS$ radius $r$. In
Poincar\'e coordinates, $ds^2=L^2(u^2(-dt^2+dx^2)+ du^2/u^2+d\O_5^2)$,
the energy scale combines with $S^5$ into
$du^2/u^2+d\O_5^2=dX^AdX^A/(L^2u^2)$, suggesting that radial breathing
translates to insertions of $SO(6)$-traces $K_{1,1,1}$ into the dual
operator. We propose that the operators ${\cal
O}_{(J_1,J_2,J_3;n_1,n_2;n_r)}$ corresponding to the giant wave-functions
with fixed breathing number $n_r$ are the components of the operator
${\cal O}_{(\D;00;J00)(n_r)}$ obtained by inserting the $SO(6)$
trace-part $K_{1,1,1}^{n_r}$ into \eq{chiralO}, i.e.
\be {\cal O}_{(\D;00;J00)(p)}={\cal O}_{A_1\dots A_J}\d^{J+2p}
X^{A_1}\cdots X^{A_J}(X^A X^A)^p\ ,\quad \D=J+2p-6Np^2/J^2,\
p=0,1,\dots\ee
when $J\sim N$.

The anomalous dimension is negative\footnote{Negative anomalous
dimensions are not unusual. Indeed there are several examples of
multi-trace operators with anomalous dimensions that are negative both
perturbatively and in the supergravity limit, where the anomalous
dimension tends to zero from below as $N\rightarrow \infty$. We thank M.
Bianchi and Y. Stanev for discussions on this point.} and independent of
the 't Hooft coupling $\l=g_{\rm YM}^2N$. However, the results are valid for
$\l\gg 1$ and there may be additional
contributions from fermions as pointed out above. From the bulk point of
view, the negative contributions can be interpreted as the binding energy
of the closed strings on the giant. This energy should depend on the bulk
coupling constant, i.e. the bulk Planck's constant $1/N^2$, but not the
masses of the individual strings, i.e. the bulk string tension $\l/R^2$.
This is in sharp contrast to the anomalous dimensions of single-trace
operators, which are positive and depend on $\l$, as expected from the
bulk picture where they represent individual closed strings.

Analogous operator constructions are relevant also for CFT duals
of electric $M2$ giants. Here one starts by building operators in
the UV from $8$ free $SU(N)$-valued $OSp(8|4)$ supersingletons,
$X^A$, $A=1,\dots,8$, and let these flow to the IR under
deformations of the free singleton theory, where they should
correspond to the giants. Hence, in the IR one should find
\be {\cal O}_{(\D;0;J000)(p)}={\cal O}_{A_1\dots A_J}\d^{J+2p}
X^{A_1}\cdots X^{A_J}(X^AX^A)^p\ ,\quad \D=J+2p-12Np^2/J^3,\
p=0,1,\dots\ee
when $J\sim N^{1/2}$.

In the case of magnetic $D3$ and $M2$ branes, the energy formula
\eq{Emagn} suggests that dual operators are built from
subdeterminants involving $P_\phi$ scalar fields and
$S_1+\cdots+S_v+2(n_r+n_\th+n_1+\cdots+n_{m-1-v})$ derivatives,
such that the operators are protected when $n_\th=0$.

\section{Conclusions}

In this paper we have shown that the $(p+1)$-dimensional field theory
of a $p$-brane in
$AdS_m\times S^n$ admits consistent KK sphere reductions on either
$S^p\subset AdS_m$, $m=p+2$, or $S^p\subset S^n$, $n=p+2$. The
resulting $(0+1)$-dimensional models are integrable, the canonical
variables separate and the quantum mechanics consists of a breathing mode
with non-trivial potential times a set of oscillators which
describe the overall transverse motion. These models contain the
previously known 1/2 supersymmetric single-spin giant gravitons,
that have one spin in $S^n$ and expand spherically in $S^n$
\cite{susskind} or $AdS_m$ \cite{myers, hashimoto}. The magnetic model also
includes a new 1/2 supersymmetric two-spin giant, that has one
extra spin in $AdS_m$ and expands in $S^n$. The BPS bounds show
that these are the only supersymmetric solutions of these
particular spherically symmetric truncation.\footnote{There are also
other supersymmetric giants in the literature, based on wrapping
$p$-branes on  supersymmetric cycles in $AdS_m\times S^n$ or
$AdS_5\times T^{1,1}$ \cite{mikhailov, beasley}.}

There are several directions in which the $(0+1)$-dimensional sigma
models should be explored. In the cases where the original $p$-brane is
supersymmetric, one should consider supersymmetric completions by
including fermions and possibly extra bosons, and in particular examine
their contributions to the $AdS$ energies. In the case of $D3$ and $M5$
branes, in analogy with sphere reductions of supergravity, we expect the
extra bosons to be embedded into the $(p+1)$-dimensional vector and
tensor fields together with certain low-lying spherical vector and tensor
harmonics on $S^3$ and $S^5$, respectively. It is worth investigating
this in detail. Having obtained the supersymmetric $(0+1)$-dimensional
sigma models, it would be interesting to examine to what extent the
salient features of the bosonic quantum mechanics prevails.

An obvious generalization is to consider the effect on the
$(0+1)$-dimensional model from $k$-fold wrapping of the $p$-brane
on $S^p$. This gives rise to $k$ copies of the quantum mechanical
system moded out by cyclic symmetry taking the $i$'th copy to the
$(i+1)$'th copy mod $k$ (this is a global reparametrization). In
the point-particle sector, the resulting giant wave-functions in
global coordinates are built from a set of copied oscillators
$X^A_i(\x)$, $\x=1,...,k$, obeying commutation rules
$[X^A_i(\x),X^B_j(\eta)]=i\delta_{\x\eta} \eta^{AB}\e_{ij}$, where
$\eta^{AB}$ has appropriate signature. The wave-functions now
involve Young-tableaux of up to $k$ rows, in rough agreement with
the proposals on the field theory side \cite{jevicki}.

Another line of generalization is to include higher modes of the
KK spectrum on $S^p$ and obtain sigma-models in dimensions between
$(0+1)$ and $(p+1)$. For example, for the $M2$ brane, we can
consistently set to zero all harmonics on $S^2$ with non-vanishing
$L_z$ eigenvalue, i.e. drop the dependence on the cyclic
coordinate, $\chi_2$ say, while keeping the full dependence on the
remaining polar coordinate, $\chi_1$ say ($0\leq \chi_1\leq \pi$).
This should lead to a non-trivial $(1+1)$-dimensional sigma-model
with generally $(\tau,\chi_1)$-dependent fields, and it would be
interesting to study whether the integrability of the
$(0+1)$-dimensional model extends to $(1+1)$ dimensions.
Similarly, the $D3$-brane on $S^3$ and the $M5$ on $S^5$ with trivial
dependence on the cyclic coordinates should give interesting
$(1+1)$ and $(2+1)$-dimensional sigma models (on $S^3$, $\chi_i$,
$i=2,3$, are cyclic and we keep $X^{m'}(\sigma^\mu)$ and
$A_i(\sigma^{\mu})$, $\s^{\mu}=(\tau,\chi_1)$; and on $S^5$,
$\chi_i$, $i=2,4,5$, are cyclic, and we keep $X^{m'}(\sigma^\mu)$
and $b_{ij}(\sigma^\mu)$, $\s^\mu=(\tau, \chi_1,\chi_3)$).

As discussed in the Introduction, electric $p$-branes in $AdS_{p+2}$ have
semi-classical scaling behavior which make them suitable probes for
examining holography at high energies. Could they also be used for
actually defining the bulk dynamics in some certain limit? Consider, for
example, the open/closed string quantum theory on a 1/2 BPS electric $D3$
giant graviton of radius $r_0$. The running string tension, which sets
the scale for massive string excitations on top of the giant ground
state, is given by $L^2T_s(r_0)\sim L^2\cosh^2r_0/\alpha'\sim
E_0\sqrt{g_s/N}\gg1$. Hence, between the ground state and the first
massive string states, there is a large number of massless open string
excitations with energy $E\sim E_0\gg1$, and $E-E_0\ll L^2T_s(r_0)$.
These are composite operators in the vector multiplet living on the
$D3$-brane (i.e. they are multi-particle states from the world-volume
point of view), which in the physical gauge describe one-particle states
in the bulk. Consider a process with ``in-state'' prepared by first
letting the breathing mode inhale until $r\gg r_0$ and then placing out
operators, carrying distinct energy and spins. We may assume the
operators to be separated initially, so that an observer in spacetime
would see localized concentrations of energy and spin densities on the
brane. During the subsequent time-evolution, the brane first exhales. For
a spacetime observer this looks like particles falling inward to a
scattering region. The brane then breathes in again and finally reaches
large size, at which point the result of the scattering can be obtained
by computing the correlator with an ``out-state''.

The question is, how good an approximation it is to describe the
whole scattering process using only the massless field theory on
the $p$-brane. Clearly, the initial excitation energies should not
be too high. However, as discussed in Section 4, the breathing may
cause the brane to implode, or at least pass through some region
of large world-volume curvature. For example, from the $D3$-brane
field theory point of view, the formation of strings may be
thought of as the $AdS$ analog of the BIon formation on a $D3$
brane in flat space \cite{BIons}. It would be interesting to
examine to what extent these stringy excitations of the $D3$-brane
may behave differently in an $AdS$ background as opposed to flat
space. A related question is whether
a rotating long string with energy-momentum propagating along the
directions of a giant $D3$-brane could be realized as a weakly
coupled state on the giant. Similar considerations could be
undertaken for membrane-like excitations of electric $M5$ branes
described by self-dual string solutions \cite{SDstring} and
rotating membranes \cite{SSholography,orbitingM2,boz}

Finally, let us point out the relation with the old ideas of $p$-branes
``at the end of the universe'' \cite{duff,duff2}. For example, the
transition from a strongly coupled string world-sheet to a weakly coupled
$D3$-brane world-volume at high energies and fixed $R^2T_s$, suggests a
similar transition at fixed energy and small $R^2T_s$, i.e. the
tensionless string limit. Indeed, for any finite $R^2T_s$, the running
string tension $R^2T_s(r_0)$ diverges in the IR. This limit, which is
most easily examined by replacing the spherical physical gauge by another
physical gauge given in Poincar\'e coordinates \cite{singularCFT}, yields
a superconformal $D3$-brane ``at the end of the universe''. This
world-volume theory (which should not be confused with the dual CFT) is
completely decoupled from string excitations, and therefore remains
weakly coupled in the tensionless string limit. Similar limits exist also
for the $M5$ and $M2$ branes. The perturbations of the conformal
$M2$/$D3$/$M5$ branes have loop expansion in inverse powers of $N$. This
suggests that a natural starting point for describing holography is to
start close to the boundary with operator insertions on conformal
$p$-branes, and then study the deformation of this system into the bulk
by switching on perturbations corresponding to the breathing mode.

The conformal $p$-branes provide a link between the original
supergravity theories in the UV region of the bulk, and higher
spin gauge theories in the IR region of the bulk. Indeed the
(unperturbed) conformal M2 brane world-volume is a free $OSp(8|4)$
supersingleton field theory, with conserved higher spin currents
in the world-volume \cite{Salam}. The conformal $D3$ and $M5$ brane
world-volume theories are $PSU(2,2|4)$ and $OSp(8^*|4)$ supersingleton
field theories with interactions stemming from the magnetic
background fluxes, and it is desirable to study their
applications for the higher spin symmetries. It would also be
interesting to examine the conformal limit of the
$Sp(2)$-covariant quantization of the point-particle sector (see
Appendix \ref{ap1}), since the $Sp(2)$-gauged version of the
$SO(m-1,2)$-covariant oscillators play a central role in
formulating the massless higher spin dynamics as formulated in
\cite{Vasiliev2003}.

In summary, giant $p$-branes in $AdS$ backgrounds deserve further study as they
have many intriguing properties both from the point of view of
holography and for the understanding of the nature of fundamental
interactions of M-theory and string theory at high energies.

\begin{acknowledgments}
E.S. and P.S. would like to thank the Feza G\"ursey Institute for
hospitality. We are grateful to I.H. Duru, J. Engquist,
K. Murakami, J. Plefka, E. Sokatchev and K. Zarembo for several
discussions, and in particular I. Bakas for pointing to
\cite{cooper} and J. Minahan for valuable suggestions on the
holographic interpretation of giants. The work of E.S. has been
supported in part by NSF Grant PHY-0314712.
\end{acknowledgments}

\appendix
\section{Quantization in Global Coordinates}
\label{ap1}

In this appendix, we briefly discuss the canonical quantization of
the global $x^A$ coordinates governed by the action \eq{adsact2}.
It is clear that $P_\Lambda=0$ is a primary constraint. The
Poisson bracket of $P_\L$ with the Hamiltonian leads to a chain of
secondary constraints, which can be summarized in terms of the
$Sp(2)$ generators as
\be L_{ij}\equiv X^A_iX_{A j}=M_{ij}\ ,\qquad P_\Lambda=0\
,\label{Sp2el}\ee
where $X^A_i=(x^A,P^A)$, $M_{11}=1$, $M_{12}=0$,
$M_{22}=2\tilde{N}^2L^4\L_0/\tilde{L}^2$. Here, $\L_0$ is the
constant that appears in \eq{Lagmult}. The four constraints are second
class.

In quantum theory, the canonical commutation relations read
\be [\hat X^A_i,\hat X^B_j]=i\e_{ij}\d^{AB}\ .\ee
The ordering ambiguity in \eq{Sp2el}  can be cured by demanding
that the operators $\hat L_{ij}=\hat X^A_{(i}\hat X_{j)A}$ generate
the $Sp(2)$ algebra. One can now impose the following
Casimir constraint:
\be \left(\hat L^{ij}\hat
L_{ij}-M^{ij}M_{ij}\right)|\Psi\rangle=0\ ,\label{sp2condel} \ee
together with $L_{11}|\Psi\rangle=P_\Lambda|\Psi\rangle=0$. Using
the oscillator algebra we find
\be \fr12 \hat L^{ij}\hat L_{ij}=\fr12 \hat J^{AB}\hat
J_{AB}+\fr14(n+1)(n-3)\ ,\ee
where $\hat J_{AB}=\e^{ij}\hat{X}_{[Ai}\hat{X}_{B]j}$ are the
generators of $SO(n+1)$. Using the expression for $M_{ij}$ given
below \eq{Sp2el} and the fact that $\Psi$ has $SO(n+1)$ highest
weight $(J0\dots 0)$, we find (note that $q_1^2=2L^2\L_0$, see
below \eq{4sphmom})
\be 1+\fr{\tilde{N}^2L^2}{\tilde{L}^2}\,q_1^2=(J+\fr{n-1}2)^2\ ,\ee
which is in agreement with \eq{Tai} for large $J$. The
wave-functions are given by the spherical harmonics
\be
\Psi_{(J00)p}(X^A,\Lambda)=\Psi_{A_1\dots A_J}X^{A_1}\cdots
X^{A_J}\ ,\label{elecwave} \ee
where $\Psi_{A_1\cdots A_J}$ is traceless and symmetric.
It would be interesting to repeat the above analysis also for the
magnetic case leading to $Sp(2)\times SO(m-1,2)$ covariant oscillators
$Y^P_i$.

\section{The Energy Spectrum of P\"oschl-Teller Potentials}
\label{ap0}

In this appendix, following \cite{cooper}, we summarize the spectrum
of  P\"oschl-Teller potentials.
A superpotential, $W(X)$ say, determines two ``partner
Hamiltonians'', $H_+=A^\dagger A$ and $H_-=AA^\dagger$, where
$A={d\over dX}+W(X)$ and $A^\dagger=-{d\over dX}+W(X)$, and the
partner potentials are given by the Riccati equations
$V_{\pm}=W^2\mp W'$. A family of superpotentials $W(a;X)$, where
$a$ denotes a set of parameters, is said to be shape invariant if
$A(a)A^\dagger(a)= A^\dagger(f(a))A(f(a))+R(a)$, i.e.
$V_-(a;X)=V_+(f(a);X)+R(a)$, where $f$ is a fixed function and
$R(a)$ is a constant (independent of $X$). The eigenvalue problem
for $H_+$ then has a generalized oscillator solution
($n=0,1,2,\dots$):
\bea \Psi^+_{(n)}(X) &=& A^\dagger(a)A^\dagger(f(a))\cdots
A^\dagger(f^{n-1}(a))\exp\left[-\int^X dY W(f^{n}(a);Y)\right],\nn\\
E^+_{(n)}&=&\sum_{k=0}^{n-1} R(f^k(a))\ .\eea
The P\"oschl-Teller Type I and II superpotentials are given by
($0\leq \th\leq \pi/(2\a)$, $r>0$)
\bea W_{I}(A,B,\a;\th)&=&A\tan \a \th -B\cot \a \th\ ,A,B>0\ ,\\
W_{II}(A,B,\a;r)&=& A\tanh \a r -B\coth \a r\ ,A>B>0,\eea
and the associated potentials, shape transformations and bound
state eigenvalues read
\bea V_{I\pm}&=& -(A+B)^2+{A(A\mp\a)\over \cos^2
\th}+{B(B\mp\a)\over \sin^2 \th}\ ,\nn\\
f_I(A,B,\a)&=&(A+\a,B+\a,\a)\ ,\qquad R_I=(A+B+2\a)^2-(A+B)^2\ ,\label{type1}\\
E^{I+}_{(n)}&=&(A+B+2n\a)^2-(A+B)^2\ ,\quad n=0,1,\dots\nn\\[5pt]
V_{II\pm}&=&
(A-B)^2-{A(A\pm \a)\over \cosh^2\a r}+{B(B\mp\a)\over \sinh^2\a r}\ ,\nn\\
f_{II}(A,B,\a)&=&(A-\a,B+\a,\a)\ ,\qquad R_{II}=(B-A)^2-(B-A+2\a)^2\
,\label{type2}\\
E^{II+}_{(n)}&=&(A-B)^2-(A-B-2n\a)^2  ,\quad
n=0,\dots,(A-B)/2.\nn\eea
The wave-functions obey Dirichlet conditions at $\th=0,\pi/(2\a)$
and $r=0$. In the Type II case there is a finite number of bound
states and then a continuum, $E^{II+}\geq (A-B)^2$. One can extend
Type I and II to $B=0$, provided the Dirichlet condition is
dropped in the case of Type II and imposed at $\th=\pm \pi/2$ in
the case of Type I.

In applying to the motion on a sphere, one has to be careful with
the Dirichlet conditions. If one switches on a spin, generated by
a vector field $V$, by imposing $V\Psi=i n \Psi$, then $\Psi$ has
to vanish at points where $V$ has zero norm. In the present
parametrization, there is a one-to-one correspondence between the
spins in the Cartan subalgebra and the cyclic coordinates. Hence,
if $\phi_i$ is an oscillatory coordinate, then both $\phi_i=0$ and
$\phi_i=\pi/2$ are vanishing points for spins in the Cartan
subalgebra, and hence all the Dirichlet conditions are globally
well-defined.

\section{Comparison with Bohr-Sommerfeld}
\label{ap2}

As we have seen above the quantum mechanical problem involving the
P\"oschl-Teller
potentials is exactly solvable. Now, we will use
Bohr-Sommerfeld approximation for Type I P\"oschl-Teller potential for
comparison. The action
integral is
\bea
I_\theta &=& 2\pi(n_\th+\fr{1}{2})\ ,\qquad n_\th=0,1,2,\dots\nn\\
&=& {L\tilde N\over \tilde L} \oint d\theta
\left[q^2-{p^2\over \cos^2 \theta}-{u^2\over \sin^2
\theta}\right]^{1/2}.\label{thetaquant}
\eea
To evaluate the integral perturbatively, we first set
$x=\sin^2\theta$ which yields
\be I_\theta = \frac{L\tilde N}{\tilde L}q \int_{b_1}^{b_2} dx
{\sqrt{f(x)} \over x(1-x)}=2\,\pi \,(n_{\theta}+\fr12) , \label{newint} \ee
where
\be f(x)=-x^2+[\fr{q^2-p^2+u^2}{q^2}]x -
\fr{u^2}{q^2}\equiv(x-b_1)(b_2-x), \label{xxx}\ee
and $0<b_1,b_2<1$. For small oscillations we need
$(b_2-b_1)\sim 0$ and $(b_1+b_2)\sim 1$, implying $p\sim u$ and
$q\sim 2u$, and we can expand in
\be \eta\equiv (q-p-u).\ee
Defining \, $2x=(b_2-b_1)y +(b_2+b_1)$ and using the approximations
\bea &&\left[1-(b_2+b_1)\right]^2\simeq \fr{(p-u)^2}{(p+u)^2}
\left[1-\fr{4}{p+u}\eta\right]\\
&&\frac{(b_2-b_1)^2}{4} \simeq \fr{2p u}{(p+u)^3}\left[\eta+
  \fr{u^2-5pu+p^2}{2pu(p+u)}\eta^2-
\fr{3u^2-10pu+3p^2}{2pu(p+u)^2}\eta^3\right] \nn \\
&&\left[(\frac{b_1+b_2}{2})(1-\frac{b_1+b_2}{2})\right]^{-1}
\simeq \frac{(p+u)^2}{pu}
\left[1-\frac{(p-u)^2}{pu(p+u)}\eta+
\fr{(p-u)^2(2p^2+pu+2u^2)}{2p^2u^2(p+u)^2}\eta^2\right]\nn
\eea
the integral (\ref{newint}) up to fourth order in $\eta$  reads (the
odd powers of $y$ do not contribute)
\be \left[2\eta -\fr{p^2-pu+u^2}{pu(p+u)}\eta^2+
  \fr{p^4-p^2u^2+u^4}{p^2u^2(p+u)^2}\eta^3+...\right]
\int_{-1}^{1} dy \sqrt{1-y^2}\left[1+h_1y^2+ h_2y^4 + ...\right]
\ee
where
\bea h_1&=&\fr{2(p^2-pu+u^2)}{pu(p+u)}\eta
-\fr{3(p^4-p^2u^2+u^4)}{p^2u^2(p+u)^2}\eta^2 \, , \\
h_2&=&\frac{4[p^4-p^3u+P^2u^2-pu^3+u^4]}
{p^2u^2(p+u)^2} \eta^2 \, .\eea
Up to this order, the integration gives
\be 2n_\th+1=\frac{L\tilde N}{\tilde L} \eta.\ee
As in the exact result \eq{Tai} there is no expansion in
$n$. However, $1/T$ corrections that are present in \eq{bspt1} and
\eq{bspt3} are not observed.

The above analysis also applies to Type II P\"oschl-Teller potential
which appeared  in \eq{prs}. One needs to introduce a new variable
$x=-\sinh^2  r$ after which the action integral becomes \eq{xxx}.

\end{document}